\documentclass[aps,prd,preprint,showpacs,showkeys,nofootinbib,superscriptaddress]{revtex4-1}

\usepackage{amsmath}
\usepackage{amssymb,slashed}
\usepackage{dcolumn}
\usepackage{bm}
\usepackage{feynmp}                                     
\usepackage{hyperref,graphicx}           
\usepackage{caption}
\usepackage{subcaption}
\usepackage{color}

\begin{document}

\title{Discovering the Twin Higgs Boson with Displaced Decays}

\author{Can Kilic}
\affiliation{Theory Group, Department of Physics\\ University of Texas at Austin,
Austin, TX 78712, USA}
\author{Saereh Najjari}
\affiliation{Theoretische Natuurkunde \& IIHE/ELEM,\\ Vrije Universiteit Brussel, Pleinlaan 2, 1050 Brussels, Belgium}
\author{Christopher B. Verhaaren}
\affiliation{Center for Quantum Mathematics and Physics (QMAP), Department of Physics,\\ University of California, Davis, CA, 95616 USA.}

\date{\today}
\begin{abstract}
The Twin Higgs mechanism keeps the scalar sector of the Standard Model (SM) natural while remaining consistent with the non-observation of new colored particles at the Large Hadron Collider (LHC). In this construction the heavy twin Higgs boson provides a portal between the SM particles and the twin sector, but is quite challenging to discover at colliders. In the Fraternal Twin Higgs setup, where light twin quarks are absent, we study a novel discovery channel for the heavy twin Higgs boson by considering its decay to a pair of light Higgs bosons, one of which subsequently decays to glueball states in the twin sector, leading to displaced vertex signatures. We estimate the sensitivity of existing LHC searches in this channel, and assess the discovery potential of the high luminosity (HL) LHC. We show that the glueballs probed by these searches are outside the sensitivity of existing searches for exotic decays of the light Higgs boson. In addition, we demonstrate that the displaced signals we consider probe a region of heavy Higgs masses beyond the reach of prompt signals. We also comment on the possibility of probing the input parameters of the microscopic physics and providing a way to test the Twin Higgs mechanism with this channel.
 \end{abstract}

\pacs{}%

\keywords{}

\maketitle
\preprint{UTTG-26-18}

\section{Introduction\label{s.intro}}
Since the discovery of the Higgs boson~\cite{Aad:2012tfa,Chatrchyan:2012ufa}, a dedicated experimental program has sprung up to determine its properties. The scalar nature of the Higgs makes it especially interesting. Considering the Standard Model (SM) as an effective field theory (EFT), one would generically expect an elementary scalar to have a mass near the cutoff of the theory, or equivalently the scale of new physics, unless protected by a symmetry~\cite{Fayet:1977yc,Dimopoulos:1981zb,ArkaniHamed:2001nc}. So far, the Large Hadron Collider (LHC) has seen no evidence of strongly coupled dynamics connected to the Higgs boson, or colored symmetry partners typically required for new symmetries to stabilize the weak scale. Precision electroweak measurements predating the LHC also imply that the scale of new physics is greater than about 5 TeV~\cite{Barbieri:2000gf}. Thus, even if new states appearing at that scale cancel the UV sensitivity to higher energies, there remains the little hierarchy problem connected to the tuning required for a light Higgs boson below the scale of new physics.

Realizations of neutral naturalness~\cite{Chacko:2005pe,Barbieri:2005ri,Chacko:2005vw,Burdman:2006tz,Poland:2008ev,Cai:2008au,Craig:2014aea,Craig:2014roa,Batell:2015aha,Katz:2016wtw,Csaki:2017jby,Serra:2017poj,Cohen:2018mgv,Cheng:2018gvu,Xu:2018ofw} provide a solution to the little hierarchy problem while explaining the lack of new particles at the LHC. These models posit symmetry partners to SM fields that are not charged under SM color, explaining null results at the LHC~\cite{Burdman:2014zta}. The prime example of such a set up is the twin Higgs~\cite{Chacko:2005pe}.

In the original twin Higgs framework, the entire SM, including its gauge groups, is duplicated in a twin sector, which is related to the visible sector by a discrete $\mathbb{Z}_2$ symmetry. In addition, the Higgs potential exhibits an approximate global symmetry, $SU(4)\times U(1)$ in the simplest versions, in which the electroweak gauge symmetries of both sectors are embedded. The SM Higgs doublet is among the pseudo-Nambu-Goldstone bosons (pNGBs) resulting from the spontaneous breaking of this global symmetry, making it naturally light. The mass of the physical Higgs boson is protected from large quantum corrections by a combination of the global symmetry and the $\mathbb{Z}_2$.   

While the initial realization of the twin Higgs mechanism faithfully doubles the SM field content, the theory can be natural with only the third generation in the twin sector. Indeed, cosmology seems to favor removing the lighter twin states and the twin photon, though mirror twin cosmologies continue to be fruitfully explored~\cite{Craig:2016lyx,Chacko:2016hvu,Barbieri:2016zxn,Chacko:2018vss}. The so-called fraternal twin Higgs (FTH) model~\cite{Craig:2015pha} realizes this more minimal setup. 

The FTH model also contains viable dark matter candidates, such as the dark baryons as asymmetric dark matter~\cite{Garcia:2015toa,Farina:2015uea}, and other possibilities~\cite{Garcia:2015loa,Hochberg:2018vdo,Cheng:2018vaj} . Cosmological issues connected to large and small scale structure~\cite{Prilepina:2016rlq} and baryogenesis~\cite{Farina:2016ndq} can also be addressed.

A less studied aspect of the twin Higgs construction is the radial mode of the Higgs sector. Like the light Higgs boson, this heavier state provides a portal between the SM and twin states. In fact, without introducing new states not dictated by the $\mathbb{Z}_2$ symmetry~\cite{Bishara:2018sgl}, or kinetic mixing between the SM and twin hypercharge, these two scalars are the only low-energy portals between the sectors, making their study particularly important. Though the light Higgs $h$ couples more strongly to visible states, the heavy Higgs $H$ is more strongly coupled to the twin states. This reduces its production cross section at the LHC, making it difficult to discover and study. The decay modes of this heavy twin Higgs were discussed in FTH-like contexts~\cite{Craig:2015pha,Craig:2016kue} and its LHC phenomenology has been recently explored~\cite{Ahmed:2017psb,Chacko:2017xpd}. These studies confirm the difficulty of using conventional collider searches to discover the heavy twin Higgs. 

Fortunately, there is an alternative search strategy. In the FTH scenario, the lightest twin hadrons are glueballs. The lightest of the glueball states has the right quantum numbers to mix with the light Higgs, inducing decays to SM states with a macroscopic decay length. The exotic decays of the light Higgs into these glueball states can be a powerful probe of the FTH scenario~\cite{Curtin:2015fna,Csaki:2015fba}. Of course, the heavy Higgs itself can decay into twin glueballs as well. While these displaced vertex (DV) signatures of the heavy Higgs have been pointed out, they are only now being systematically explored~\cite{Alipour-fard:2018mre}.

The glueballs, if discovered, would also offer valuable insight into the nature of the twin sector. Their mass and decay length can yield information about the running of the twin QCD coupling as well as the magnitude of the coupling between the light Higgs and the twin quarks, potentially offering a consistency check of the ${\mathbb Z}_2$ symmetry structure of the underlying theory. Combined with an observation of the signal cross section, which contains additional information about the coupling between the light and heavy Higgs bosons, the consistency check can be extended to the full symmetry structure of the twin Higgs framework.

In this work we show how displaced decays can be used to discover the heavy twin Higgs at the LHC. We begin in Sec.~\ref{s.HiggsPhys} by reviewing the salient features of the Twin Higgs scenario. Section~\ref{s.TwinGlue} then describes the the properties of twin glueballs in the FTH framework. In Sec.~\ref{s.exp} we study the reach of existing LHC searches for the glueballs, and we assess the future discovery potential in this channels. In particular, while for glueballs with shorter decay length the production of either Higgs boson with a decay directly to glueballs appears challenging to discover, we demonstrate that the process $H\to hh\to bb G_0G_0$ has better prospects and can be used to discover the heavy twin Higgs at the LHC. We also comment on how this channel may offer a way to explore the structure of the twin sector. We conclude in Sec.~\ref{s.con}.

\section{Overview of the Twin Higgs Scenario\label{s.HiggsPhys}}
In this section we review the salient aspects of the FTH framework, leaving additional details to Appendix~\ref{a.Model}. The scalar sector contains a fundamental of $SU(4)$, $\mathcal{H}$, which is composed of two doublets $H_A$ and $H_B$ under the gauged $SU(2)_A\times SU(2)_B$ subgroup, where we associate the label $A$ with the SM sector and $B$ with the twin sector. As mentioned in the introduction, this symmetry structure includes the SM Higgs boson $h$ as a pNGB once the $SU(4)$ is spontaneously broken.

Since our focus is on the production and decay of the light and heavy Higgs bosons at the LHC, we consider their interactions in detail. Adopting the notation of ref.~\cite{Barbieri:2005ri}, the scalar potential can be written as
\begin{align}
V=&-\mu^2\left(H_A^\dag H_A+H_B^\dag H_B \right)+\lambda\left(H_A^\dag H_A+H_B^\dag H_B \right)^2\nonumber\\
&+m^2\left( H_A^\dag H_A-H_B^\dag H_B\right)+\delta\left[\left( H_A^\dag H_A\right)^2+\left(H_B^\dag H_B \right)^2 \right].\label{e.HiggsPotential}
\end{align}
The $m^2$ and $\delta$ terms break the global $SU(4)$ symmetry of the Higgs sector, so they must be significantly smaller than $\mu^2$ and $\lambda$ respectively for the Higgs to be a pNGB. The radiative generation of these terms by gauge and Yukawa interactions is small enough for this purpose.

It is also useful to introduce a nonlinear parameterization of the global symmetry~\cite{Chacko:2017xpd}, including the radial mode $\sigma$,
\begin{equation}
\mathcal{H} =\left( \begin{array}{c}
H_A\\
H_B
\end{array}\right)=
\exp\left(\frac{i}{f}\Pi \right)\left(\begin{array}{c}
0\\
0\\
0\\
f+\frac{\sigma}{\sqrt{2}}
\end{array} \right) \; ,
 \end{equation}
 with the vacuum expectation value (VEV) $f$ spontaneously breaking the $SU(4)$ symmetry down to $SU(3)$. In unitary 
gauge, where all the $B$ sector NGBs have been eaten by the corresponding vector bosons, we have
 \begin{equation}
\Pi=\left(\begin{array}{ccc|c}
0&0&0&ih_1\\
0&0&0&ih_2\\
0&0&0&0\\ \hline
-ih_1^{\ast}&-ih_2^{\ast}&0&0
\end{array} \right).
 \end{equation}
After going to unitary gauge in the $A$ sector as well, $h_1=0$ and $h_2=(v+\rho)/\sqrt{2}$, which leads to
 \begin{align}
\displaystyle H_A&= \left( \begin{array}{c}
0\\
\left(f+\frac{\sigma}{\sqrt{2}} \right)\sin\left( \frac{v+\rho}{\sqrt{2}f} \right)
\end{array}\right), &
 H_B&= \left(\begin{array}{c}
0\\
\left(f+\frac{\sigma}{\sqrt{2}} \right)\cos\left( \frac{v+\rho}{\sqrt{2}f}  \right)
\end{array} \right).
\end{align}

We label the VEVs of the two fields 
\begin{equation}
v_A=v_\text{EW}=\sqrt{2}f\sin\vartheta , \ \ \ \ v_B=v_\text{Twin}=\sqrt{2}f\cos\vartheta,
\end{equation}
with $\vartheta\equiv v/(f\sqrt{2})$. It follows that the vector and fermion particle masses in each sector satisfy
\begin{equation}
m_\text{Twin}=m_\text{SM}\cot\vartheta \, .\label{e.massRelation}
\end{equation}
The scalar masses do not follow this simple relation. The scalar potential in Eq.~\eqref{e.HiggsPotential} leads to mass mixing between the physical pNGB $\rho$ and the radial mode $\sigma$. We see this by rewriting the potential as
\begin{align}
V=&f^2\left(1+\frac{\sigma}{\sqrt{2}f}\right)^2\left[ -\mu^2-m^2\cos\left( \frac{\sqrt{2}(v+\rho)}{f}\right)\right]\nonumber\\
&+f^4\left(1+\frac{\sigma}{\sqrt{2}f}\right)^4\left[\lambda+\delta -\frac{\delta}{2}\sin^2\left( \frac{\sqrt{2}(v+\rho)}{f}\right) \right].
\end{align}
Demanding that the one-point functions of the two physical scalars vanish leads to the conditions
\begin{align}
\mu^2+m^2\cos(2\vartheta)&=f^2\left[2(\lambda+\delta)-\delta\sin^2(2\vartheta) \right],\\
m^2&=f^2\delta \cos(2\vartheta).
\end{align}
This allows us to express $\mu^{2}$ as
\begin{equation}
\mu^2=f^2\left(2\lambda+\delta \right),
\end{equation}
and the mass mixing matrix between $\sigma$ and $\rho$ can be written as
\begin{equation}
\left(\begin{array}{cc}
\sigma & \rho
\end{array}\right)f^2\left(\begin{array}{cc}
4(\lambda+\delta)-2\delta\sin^2(2\vartheta) & -\delta\sin(4\vartheta)\\
-\delta\sin(4\vartheta) & 2\delta \sin^2(2\vartheta)
\end{array} \right)\left( \begin{array}{c}
\sigma\\
\rho
\end{array}\right).\label{e.massMix}
\end{equation}
Finally, we define the mass eigenstates by
\begin{equation}
\left(\begin{array}{c}
h\\
H
\end{array} \right)=\left( \begin{array}{cc}
\cos\theta & \sin\theta\\
-\sin\theta & \cos\theta
\end{array}\right)
\left(\begin{array}{c}
\rho\\
\sigma
\end{array} \right),
\end{equation}
where
\begin{equation}
\begin{array}{cc}
\displaystyle \sin(2\theta)=\frac{2f^2\delta\sin(4\vartheta)}{m_H^2-m_h^2}, & \displaystyle \cos(2\theta)=\frac{4f^2\left[\lambda+\delta\cos^2(2\vartheta)\right]}{m_H^2-m_h^2},
\end{array}
\end{equation}
and
\begin{equation}
m_{H,h}^2=2f^2\left[ \lambda+\delta\pm\sqrt{\lambda^2+\delta(2\lambda+\delta)\cos^2(2\vartheta)}\right] .
\end{equation}
We refer to $h$ and $H$ as the light and heavy Higgs boson, respectively. The scalar potential guarantees that the masses of the eigenstates satisfy the relation
\begin{equation}
m_H\geq m_h\cot\vartheta=m_h\frac{m_T}{m_t},\label{e.HmassBound}
\end{equation}
where $m_T$ and $m_t$ correspond to the masses of the twin and SM top quark, respectively.  In the Appendix we show that the couplings of the light Higgs to SM particles has a factor of $\cos(\vartheta-\theta)$ compared to that of the SM Higgs boson, while the couplings of the heavy Higgs to SM particles has a factor of $\sin(\vartheta-\theta)$. Similarly, the couplings of $h$ to twin particles are those of a SM Higgs boson with $v_\text{EW}\to v_\text{Twin}$ multiplied by $\sin(\vartheta-\theta)$, while the couplings of $H$ to twin states are those of a SM Higgs boson with $v_\text{EW}\to v_\text{Twin}$ multiplied by $\cos(\vartheta-\theta)$.

Because the couplings of the light Higgs and twin Higgs are related to the SM Higgs couplings by simple factors we can immediately write down their production cross sections in terms of the SM predictions:
\begin{align}
\sigma(pp\to h)&=\cos^2(\vartheta-\theta)\sigma(pp\to h(m_h))_{\text{SM}},\nonumber\\
\sigma(pp\to H)&=\sin^2(\vartheta-\theta)\sigma(pp\to h(m_H))_{\text{SM}}.\label{e.production}
\end{align}
Here $\sigma(pp\to h(m))_{\text{SM}}$ denotes the SM Higgs production cross section for a Higgs boson of mass $m$. The LHC Higgs Cross Section Working Group has calculated precise values for the SM Higgs boson production cross section as a function of its mass~\cite{deFlorian:2016spz}. We obtain the production cross sections for either $h$ or $H$ at the LHC by modifying these results according to Eq. \eqref{e.production}.

The $\cos(\vartheta-\theta)$ factor leads to deviations in the observed rates of the SM-like Higgs boson~\cite{Burdman:2014zta}. However, in the limit $m_H \to m_h\cot\vartheta$, $\theta$ becomes equal to $\vartheta$, eliminating the deviations. Unfortunately, in the same limit the fine-tuning in the model increases and becomes comparable to that in the SM unless $m_H>m_T$~\cite{Chacko:2017xpd}. 

\begin{figure}
\begin{centering}
\includegraphics[width=0.75\textwidth]{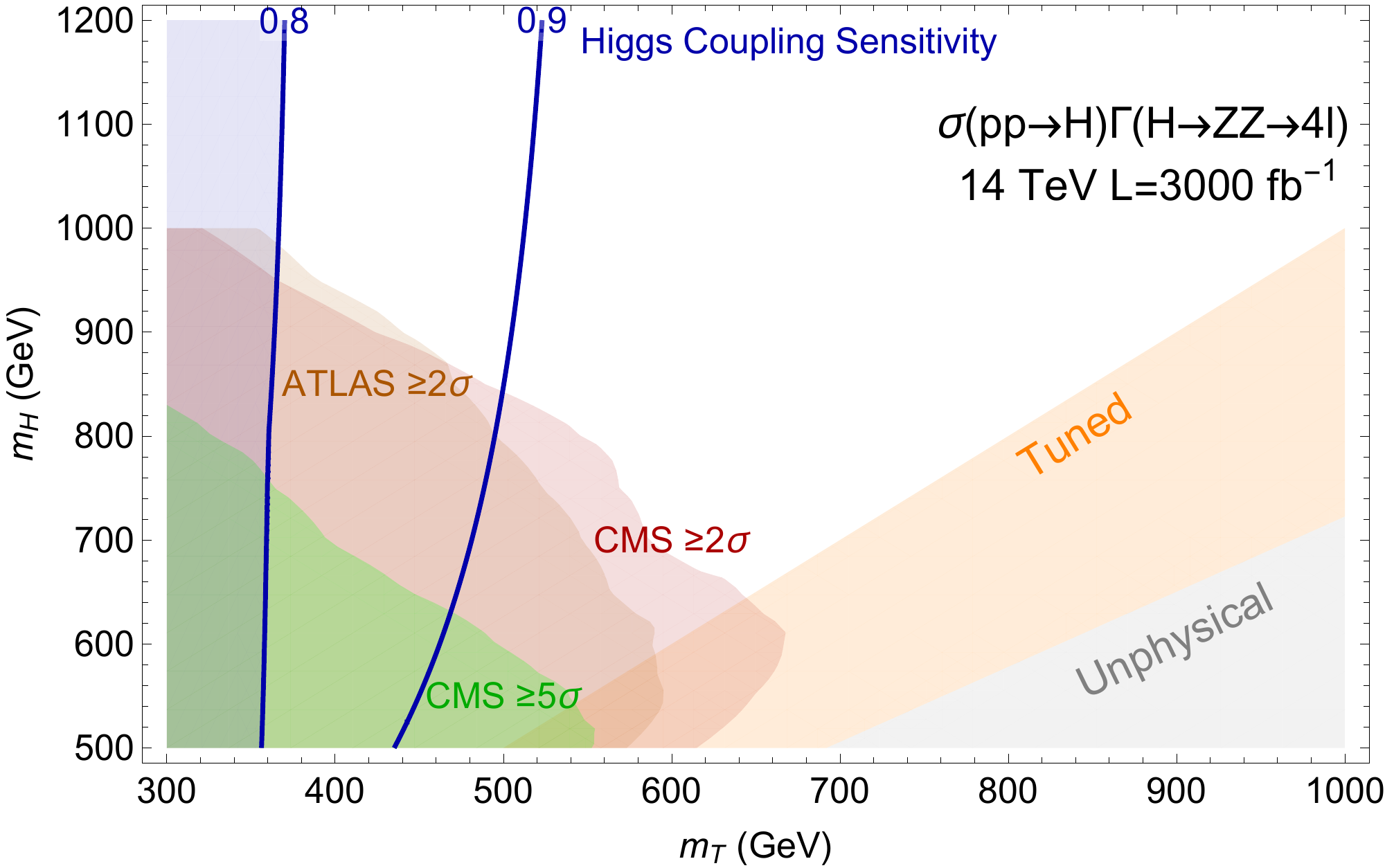}
\end{centering}
\caption{\label{f.HiggsCouplings} Reach of HL-LHC from heavy Higgs to $ZZ\to4\ell$ shown as the red and green shaded regions. The blue contours indicate the ratio of $(\sigma\times{\rm BR})$ of the light Higgs to that of the SM Higgs. In the orange region, the fine-tuning is not significantly improved relative to the SM. The gray shaded region is theoretically inaccessible due to the lower bound on the mass of the heavy Higgs.}
\end{figure}

We show the ratio of $(\sigma\times{\rm BR})$ in the FTH model to the SM prediction
\begin{equation}
\frac{\sigma(pp\to h)_\text{FTH}\times\text{BR}(h\to \text{SM})_\text{FTH}}{\sigma(pp\to h)_\text{SM}\times\text{BR}(h\to \text{SM})_\text{SM}},
\end{equation}
as a function of $m_H$ and $m_T$ in Fig.~\ref{f.HiggsCouplings} as the blue contours. LHC measurements have already determined the SM-like Higgs boson rates to be within 20\% of the SM prediction~\cite{Khachatryan:2016vau}, and therefore the area to the left of the 80\% contour in Fig.~\ref{f.HiggsCouplings} is excluded. By the end of it high luminosity run, the LHC is projected to probe deviations at the 10\% level~\cite{Dawson:2013bba}. 

As we can see in Fig.~\ref{f.HiggsCouplings}, Higgs coupling measurements favor $m_T\geq 350$ GeV. The bound in Eq.~\eqref{e.HmassBound} requires $m_H\geq 250$ GeV, but the tuning bound $m_H>m_T$ is more constraining. Together these imply that the twin Higgs is heavy enough to decay to real pairs of SM top quarks, the light Higgs, and weak vector bosons, which dominate the visible decay width. Both ATLAS~\cite{ATL-PHYS-PUB-2013-016} and CMS~\cite{CMS-PAS-FTR-13-024} have made high luminosity (HL-LHC) projections for discovering a heavy scalar in the $H\to ZZ\to 4\ell$ channel, see also~\cite{Holzner:2014qqs}. Both the exclusion and discovery projections can be translated into the FTH parameter space. We do this by modifying the Higgs cross section working group production cross sections according to Eq.~\eqref{e.production}, and multiplying by the branching fraction $\text{BR}(H\to ZZ)$ in the FTH model, comparing the result with the experimental limits. The resulting CMS discovery reach (green) and the exclusion reach of both experiments (red) are shown in Fig.~\ref{f.HiggsCouplings}. While discovery is possible for lighter $m_H$, reaching to higher (somewhat more natural) heavy Higgs masses requires different search strategies.

In the rest of this paper we study the rare but distinctive decay channel of the light and heavy Higgs bosons to glueball states. As shown in~\cite{Juknevich:2009gg}, the lightest glueball $G_0$ has a small mixing with the light Higgs, leading to displaced decays. The heavy Higgs also couples to hidden glueballs, and with a larger effective coupling since the heavy Higgs is mostly a twin sector particle. However, because the pNGBs dominate the width of $H$, the branching fraction to twin glueballs is suppressed, $\text{BR}(H\to g_Bg_B)\sim 10^{-4}$. This, along with the reduced production cross section, makes it difficult to discover the heavy Higgs through direct decays to glueballs.

Fortunately, there is another path from the twin Higgs to glueballs: the light Higgs. As long as they are all kinematically accessible, the heavy Higgs decays equally to each of the pNGBs of the broken global symmetry, the EW gauge bosons of each sector and the light Higgs. These decays dominate the $H$ width, giving each pNGB a branching fraction of about $1/7$. For smaller $m_H$, kinematic effects reduce the branching to the heavier twin states, increasing the light Higgs rate, so in general $\text{BR}(H\to hh)\gtrsim 1/7$. If one of these light Higgs bosons subsequently decays into glueballs, the final state contains a DV recoiling against the SM-like decay products of the second light Higgs, which facilitates triggering. When the top partner mass is not too heavy, this channel produces many more glueballs than direct decays of the heavy twin Higgs to twin glue.

\section{Properties of Twin Glueballs\label{s.TwinGlue}}
Having reviewed the scalars in the Twin Higgs setup, we turn our attention to the twin glueballs, the lightest hadrons in the twin sector when the twin quarks are heavy. The spectrum of pure $SU(3)$ gauge theories has been computed on the lattice \cite{Morningstar:1999rf,Chen:2005mg}. There are 12 stable  $J^{PC}$ eigenstates in the glueball spectrum, the lightest of which is the $0^{++}$ state. This state, which we refer to as $G_0$, has mass $m_0$, while the heaviest states have masses $\lesssim3m_0$. The glueball spectrum is directly tied to the strong scale of twin QCD, with $m_0 \approx 6.8 \Lambda_{\text{QCD}}$ in the $\overline{\mathrm{MS}}$ renormalization scheme. Assuming that the $\mathbb{Z}_2$ symmetry forces the SM and twin QCD couplings to be equal at the UV cutoff scale, $\Lambda_{\text{QCD}}$, and thereby the glueball mass spectrum, is determined from the running of the twin strong coupling constant and is a function of the twin quark masses. In the FTH set-up, the range of interest for $m_{0}$ is found to be $15-30$ GeV~\cite{Curtin:2015fna}.

In~\cite{Juknevich:2009gg} the decays of the glueballs of a hidden QCD sector were computed. The starting point of the analysis is the effective coupling of the Higgs to the hidden gluon field strengths
\begin{align}
\delta \mathcal{L}^{(6)} &= \frac{\alpha_s^\text{B}}{3 \pi} \left[ \frac{y^2}{M^2} \right]   |H|^2\text{Tr}\,  {G}_{\mu \nu}  {G}^{\mu\nu}\nonumber\\
&\supset \frac{\alpha_s^\text{B}}{6 \pi} \left[ \frac{y^2}{M^2} \right] v_\text{EW}h \,{G}^{a}_{\mu \nu}  {{G}^{a}}^{\mu\nu}.\label{e.dimsix}
 \end{align}
Here $\alpha_s^\text{B}$ is the strong coupling constant in the hidden sector and in the second line we have set $H \to (0, (v_\text{EW}+h)/\sqrt{2})^T$. The factor $\left[ y^2 / M^2 \right]$ is model-dependent, and we now calculate it for the FTH setup.

For this, we can apply the Higgs low-energy theorems~\cite{Ellis:1975ap,Shifman:1979eb,Kniehl:1995tn} to the Twin Higgs construction. In particular, the operator of interest is contained in~\cite{Gillioz:2012se}
\begin{equation}
\frac{\alpha_s^\text{B}}{16\pi}{G}^{a(\text{B})}_{\mu \nu}  {{G}^{a(\text{B})}}^{\mu\nu}\sum_i \delta b_i\ln m_i^2(h),
\end{equation}
where the sum is over particles in the fundamental representation of $SU(3)_B$ and $\delta b_i$ is 2/3 for a Dirac fermion. Note that, in the SM, $m_t(h)=m_t(1+h/v_\text{EW})$, which leads to the well-known coefficient of the operator~\cite{Carmi:2012yp}
\begin{equation}
\frac{\alpha_s}{24\pi}\frac{\lambda_t^2}{m_t^2}\left|H \right|^2{G}^{a}_{\mu \nu}  {{G}^{a}}^{\mu\nu}.\label{e.SMHiggsGlue}
\end{equation}
From Eq.~\eqref{e.fermionCoupling}, we find the equivalent twin Higgs result 
\begin{equation}
m_T(h)=m_T\left[1-\frac{h}{v_\text{Twin}}\sin(\vartheta-\theta)\right].
\end{equation}
which leads to 
\begin{equation}
\left[ \frac{y^2}{M^2}\right]=\frac{\tan\vartheta}{2v_\text{EW}^2}\sin(\vartheta-\theta).\label{e.yOverM}
\end{equation}

The dimension-6 operator in Eq. \eqref{e.dimsix} allows glueballs to decay to SM particles through an off-shell Higgs. The decay width for $G_0$  decaying to two SM particles is given by~\cite{Juknevich:2009gg}
\begin{equation}
\Gamma(G_0 \to \text{SM})  \ \ = \ \  
\left( \frac{1}{12 \pi^2} \left[\frac{y^2}{M^2} \right]\frac{v_\text{EW} }{m_h^2 - m_0^2} \right)^2 
\
\left(4 \pi \alpha_s^\text{B} \mathbf{F_{0^{++}}^S}\right)^2
\ \ 
\Gamma^\mathrm{SM}_{h\to \text{SM}}(m_0^2),\label{e.glueballwidth}
\end{equation}
where $4 \pi \alpha_s^\text{B} \mathbf{F_{0^{++}}^S}=4 \pi \alpha_s^\text{B}\langle 0|\text{Tr}\ G_{\mu\nu}^\text{(B)}G^{\text{(B)}\mu\nu}|0^{++}\rangle\approx 2.3 m_0^3$~\cite{Curtin:2015fna} and $\Gamma^\mathrm{SM}_{h\to \text{SM}}(m_0^2)$ is the decay width of a SM Higgs with mass $m_0$. Consequently, $G_0$ has exactly the branching fractions of a SM Higgs with mass $m_0$. Also, because $\Gamma^\mathrm{SM}_{h\to \text{SM}}(m_0^2)\sim m_0$ the total $G_0$ width scales as $m_0^7$. Based on Eq.~\eqref{e.yOverM} the width also scales like $\sin^4\vartheta$. This means that the glueball decay lengths are quite sensitive to the glueball mass and the twin top mass. 

\begin{figure}
\begin{centering}
\includegraphics[width=0.8\textwidth]{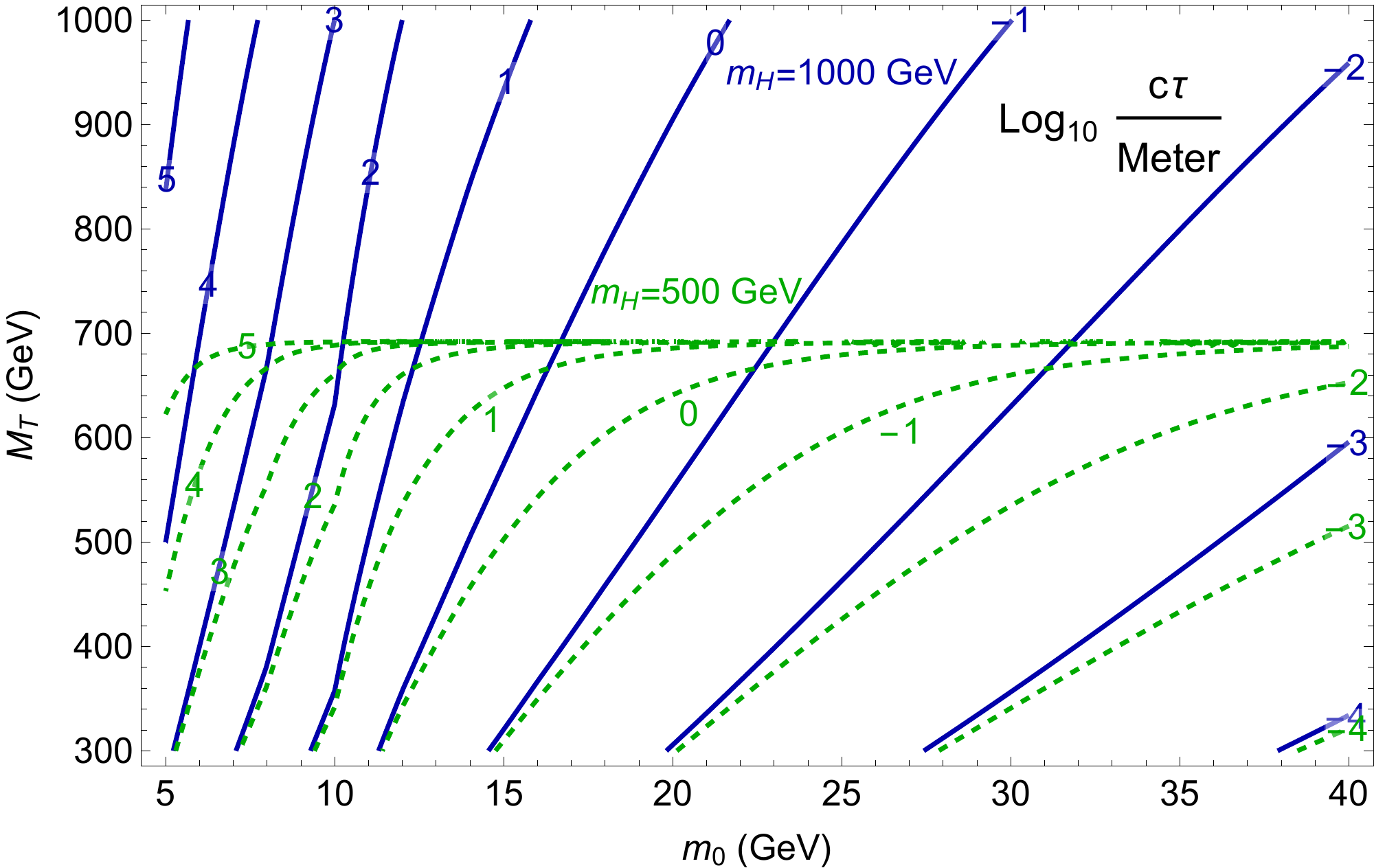}
\end{centering}
\caption{\label{f.lifetime}Contours of the $\log_{10}\frac{c\tau}{\text{meter}}$ as a function of the glueball mass $m_0$ and the twin top-quark mass $m_T$. The solid (dashed) contours take the mass of the heavy Higgs to be 1000 (500) GeV. Near the lower bound of the heavy Higgs mass the decay lengths become arbitrarily long.}
\end{figure}

The $G_0$ lifetime as a function of glueball mass and twin top mass is shown in Fig. \ref{f.lifetime}. We use {\sc{HDECAY}}~\cite{Djouadi:1997yw} to calculate the SM Higgs width for $m_{h}<125$~GeV. Note that near the lower bound on the heavy Higgs mass the decay lengths become much longer as $\sin(\vartheta-\theta)$ becomes smaller. However, this is not in the region of parameter space that is of interest, since it corresponds to $m_H<m_T$, where the theory is fine-tuned. 

The $G_0$ lifetime is typically much shorter than the other glueball states, though the nature of states near the cutoff~\cite{Garcia:2015loa} can allow $0^{-+}$ to decay on collider timescales. In this paper, we take a conservative approach and we only consider DV signals arising from $G_0$ production from heavy and light Higgs decays.

In the case of the light Higgs, we assume that due to kinematics no more than two glueballs are produced. Once again this is a conservative assumption since we are leaving out events with a potentially higher glueball multiplicity, and where each glueball has a smaller boost, and is therefore more likely to decay in the inner part of the detector. With this assumption we have
\begin{equation}
\label{e.brhglueballs}
\text{BR}\left(h\to G_0G_0\right)  \approx  \text{BR}(h\to g_B g_B)  \sqrt{1 - \frac{4 m_0^2}{m_h^2}} \ \cdot \  \kappa_h,
\end{equation}
where $g_B$ represents the twin gluon and $\kappa_h$ is a multiplicative factor specified by nonperturbative physics. It accounts for how often the produced glueballs are pairs of $G_{0}$, as opposed to the heavier states. We discuss the range of interest for $\kappa_{h}$ below.

We cannot make similar assumptions about the direct decays of the heavy Higgs into glueballs. Since $m_H\gg m_0$ we expect two glueball jets rather than a simple two-body decay. A simple estimate of the number of glueballs produced, as in~\cite{Chacko:2015fbc}, can be obtained by including a factor.  
\begin{equation}
\langle n(E_{\text{CM}}^2)\rangle \propto \exp\left[\frac{12\pi}{33}\sqrt{\frac{6}{\pi\alpha_s^B(E_\text{CM}^2)}}+\frac14 \ln\alpha_s^B(E_\text{CM}^2) \right],\label{e.HadMult}
\end{equation}
which is the multiplicity of any given hadron in the massless limit. While this corrects for the glueball multiplicity, it does not produce the kinematic distribution for the glueballs needed for a collider study. To account for both multiplicity and kinematics we employ a pure glue parton shower~\cite{CCV} to produce the plausible kinematics of these glueballs\footnote{See also \cite{Lichtenstein:2018kno} for a different approach in studying the glueball multiplicity in the final state.}. This shower automatically produces the multiplicity scaling of Eq.\eqref{e.HadMult}. However, we are still left with a nonperturbative input $\kappa_H$ about the relative abundance of $0^{++}$ glueballs at the end of the shower.

One way to estimate these glueball abundances is by using a thermal partition functions with $T \sim \Lambda_\mathrm{QCD'}$~\cite{Juknevich:2010rhj}. Since the glueball masses are almost an order of magnitude above the confinement scale, Boltzmann suppression significantly favors the lightest state $G_0$, despite the small mass differences. This thermal argument implies $\kappa_H \sim 0.5$, but is likely to change at lower $H$ masses. Furthermore, for glueball production in exotic Higgs decays, less energy is available so $\kappa_h$ may well be much greater than $0.5$. Indeed, some or all of the heavier glueball final states are forbidden if $m_0 \gtrsim 20$ GeV. For this work, we simply assume $\kappa_H=0.5$ as a benchmark.

So far we have neglected completely how twin quarks can affect the twin hadron spectrum. The variety of their effects is outlined in \cite{Craig:2015pha} and further explored in \cite{Cheng:2015buv}. In short, depending on the values of $\Lambda_{\text{QCD}'}$ and the twin bottom Yukawa coupling, bound states of twin bottom quarks can play a significant role in the decays of the twin Higgs. To avoid these complications, we make the simplifying assumption that the twin bottom Yukawa is large enough to ensure that $G_0$ is the lightest hadron.

\section{Existing Constraints and Future Experimental Prospects}\label{s.exp}

In this section, we review existing constraints on the Twin Higgs setup from DV-based searches, starting with light Higgs production and decay, and then processes involving the heavy Higgs. We evaluate the discovery prospects for the $H\to hh\to$~DV+X channel, and how a discovery in this channel may provide valuable information about the underlying physics.

\subsection{Light Higgs Decays\label{ss.lightHiggstToGlue}}
\begin{figure}
\begin{centering}
\includegraphics[width=0.49\textwidth]{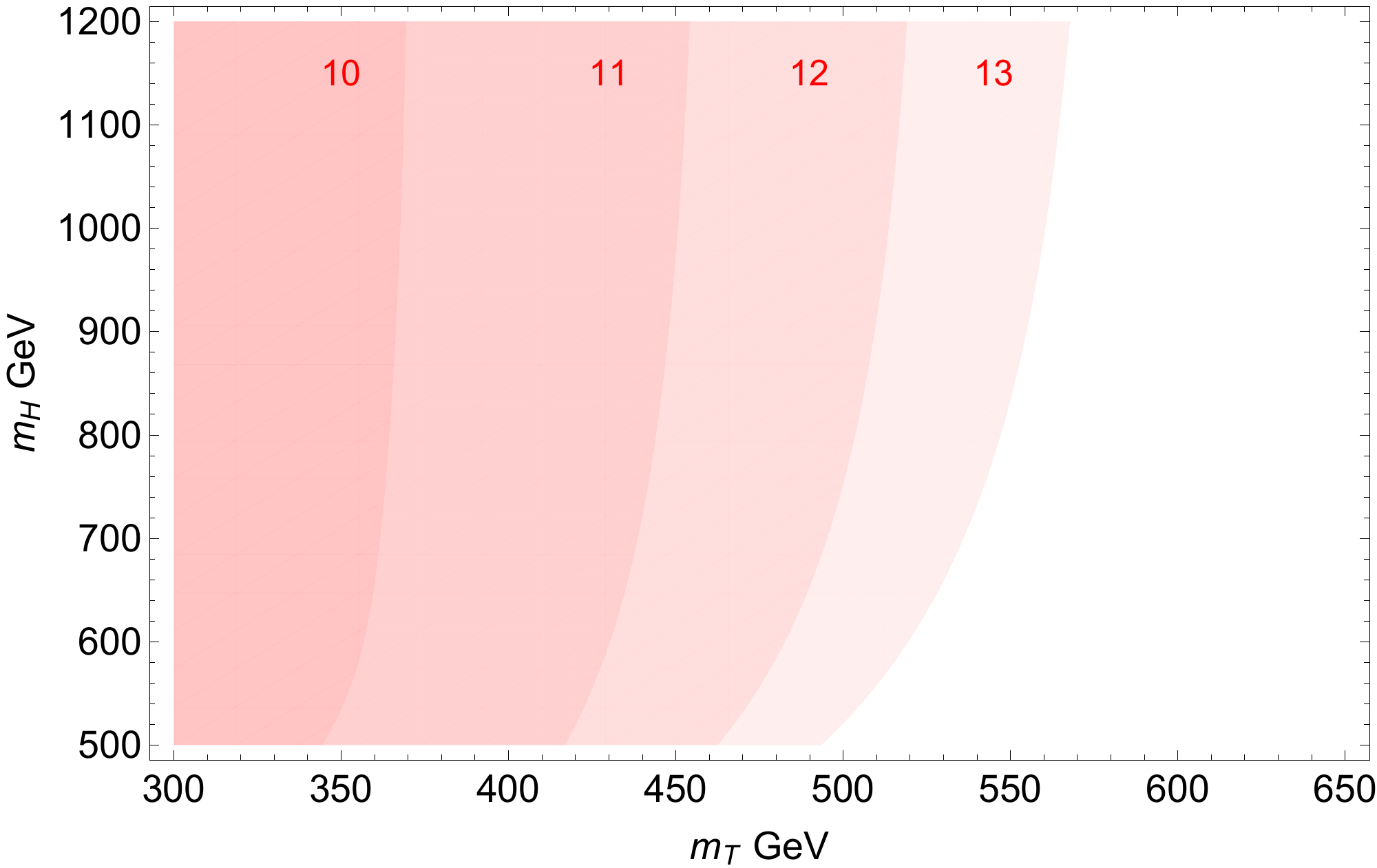}
\includegraphics[width=0.49\textwidth]{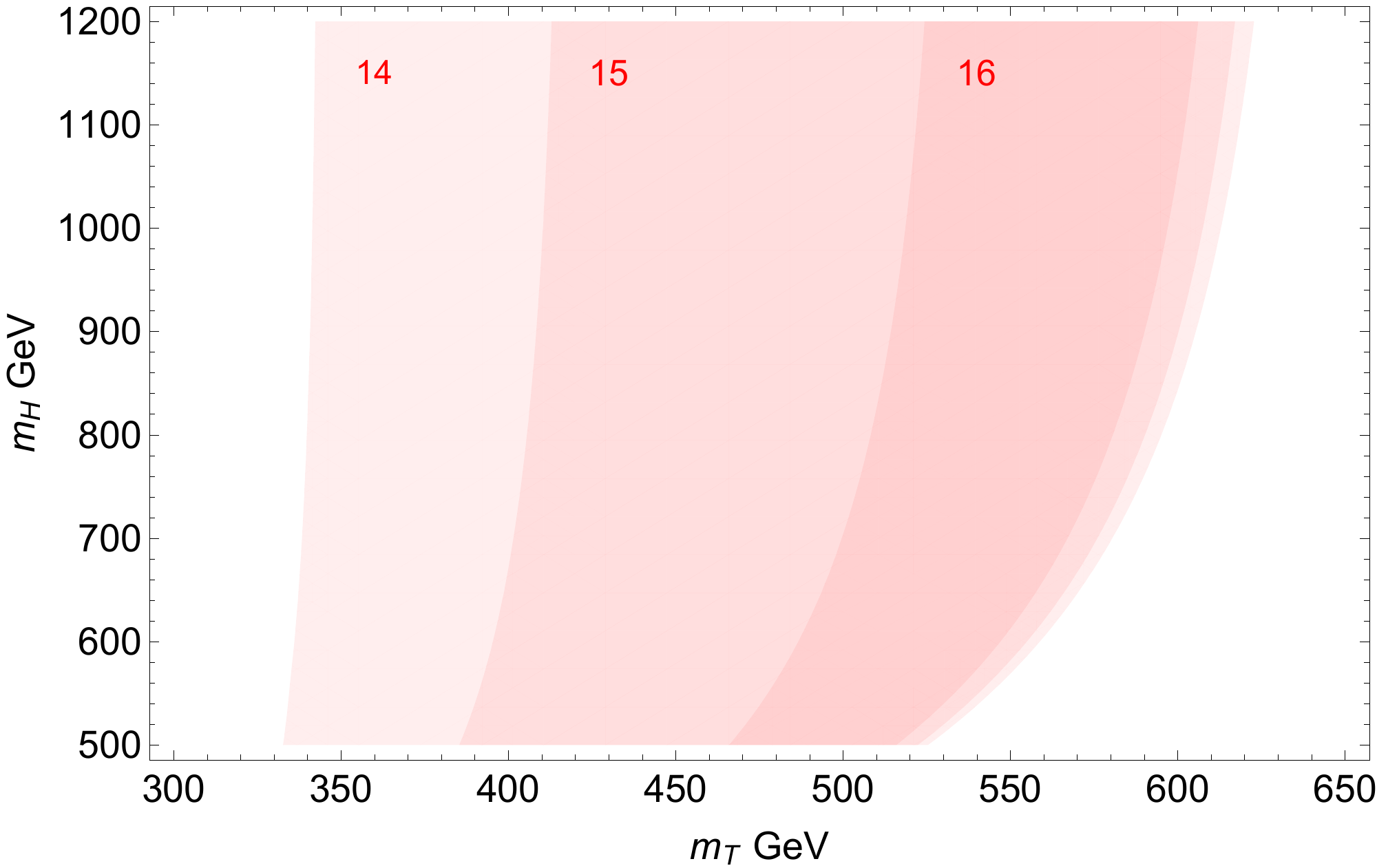}
\end{centering}
\caption{\label{f.glueExclude}  Region of parameter space excluded by ATLAS exotic Higgs decay DV search~\cite{Aaboud:2018aqj} for glueball masses from 10 to 13 GeV on left and 14 to 16 GeV on the right.}
\end{figure}

For light long-lived particles produced in the decays of the light Higgs, the muon system is a promising place to look for decays, and existing searches (see Fig.~11 of~\cite{Aaboud:2018aqj}) place constraints on $c\tau$ as a function of mass that are relevant for the glueballs in the FTH setup. In Fig.~\ref{f.glueExclude} we translate the experimental limits of this analysis into bounds on the FTH parameter space. The experimental results are only given for a few mass benchmarks, so we are forced to make certain approximations. For the masses plotted in Fig.~\ref{f.glueExclude}, we use the experimental exclusion curve for the 15 GeV scalar and as a result, the exclusion regions for $m_{0}<15$~GeV are somewhat aggressive. For each point in parameter space the Higgs branching ratio into glueballs and the glueball decay length are calculated, and compared with the experimental exclusion curve. Glueballs with mass below 10 GeV have decay lengths that are too long for to be excluded, while those with masses above 16 GeV have decay lengths that are too short to produce enough events in the muon system. 

\begin{figure}
\begin{centering}
\includegraphics[width=0.75\textwidth]{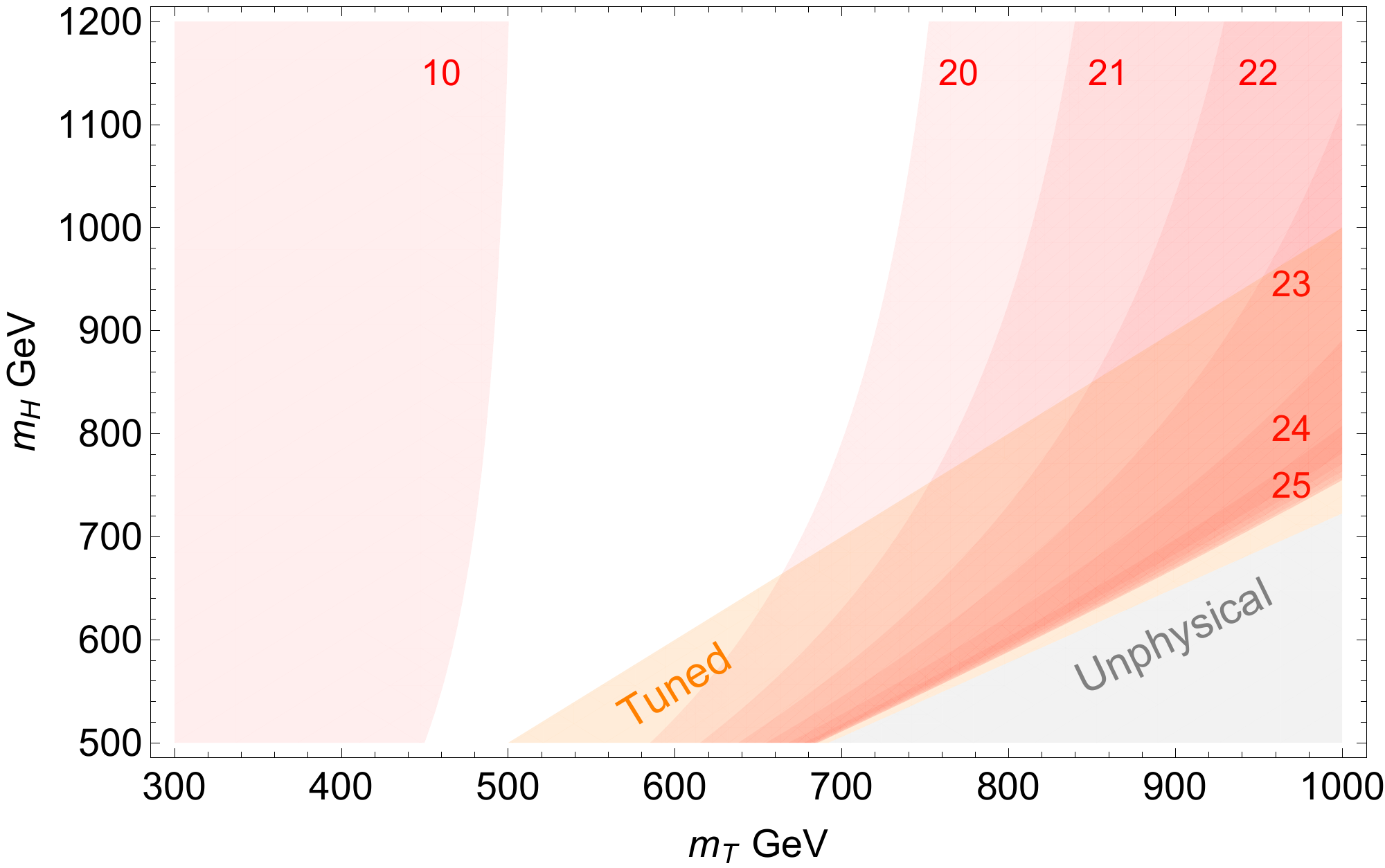}
\end{centering}
\caption{\label{f.glueExtrap} Projected region of parameter space excluded by muon system DV searches~\cite{Coccaro:2016lnz} from Higgs decays to glueballs of mass 10~GeV as well as masses in the [20,25]~GeV range. }
\end{figure}

With the added luminosity of the remaining LHC runs, the sensitivity to displaced exotic Higgs decays will increase. In ref.~\cite{Coccaro:2016lnz} these muon system searches were extrapolated through the high luminosity run (HL-LHC). In Fig.~\ref{f.glueExtrap} we apply these extrapolations to the FTH parameter space using the same procedure as in Fig~\ref{f.glueExclude}. Once again, since only benchmark masses of 10, 25, and 40 GeV were used by the experimental analysis, we approximate the sensitivity to intermediate glueball masses by using the experimental results of the closest mass benchmark. We find that the heavier glueballs are not expected to be constrained for low $m_T$. As we will soon demonstrate, these regions of lighter $m_T$ and heavier $m_0$ can be probed by relying on the heavy Higgs. 

In ref.~\cite{Curtin:2015fna}, LHC projections of displaced Higgs decays were made for other search strategies. These searches use the tracker rather than the muon system, providing access to shorter decay lengths, and therefore heavier glueballs; see also~\cite{Csaki:2015fba}. While the projected sensitivity appears promising, these types of searches have not yet been implemented experimentally, and no detailed estimation of backgrounds is currently available. Of course, the introduction of new instrumentation and search strategies, such as~\cite{Liu:2018wte} may well yield even stronger results than these extrapolations suggest. The sensitivity of future lepton colliders have also been shown to have great potential to discover these displaced Higgs decays~\cite{Alipour-Fard:2018lsf}.

\subsection{Heavy Higgs to Glueballs}
We start by considering the direct decay of the heavy Higgs to twin gluons. This signal event rate depends strongly on branching fraction $\text{BR}(H\to g_Bg_B)\sim 10^{-4}$. While this number is small, it is off-set somewhat by the number of glueballs that are produced as a result of the twin QCD-shower and hadronization into glueball states, including $G_0$ which results in DV signatures observable at the LHC. 

Triggering is the greatest challenge for this channel. The heavy Higgs is produced primarily through gluon fusion, which does not provide additional objects in the event to trigger on. Relying on hard jets from initial state radiation, or the VBF production channel reduces the signal rate to a level that makes discovery impossible. Dedicated DV triggers seem the best hope for this decay mode. Note the significant contrast between the heavy and light Higgs cases. The number of light Higgses produced is large enough that VBF and associated production triggers can still provide meaningful sensitivity into the parameter region of interest.

ATLAS has employed such dedicated triggers in the HCAL~\cite{Aad:2015asa,ATLAS:2016olj} and in the muon system~\cite{Aad:2015uaa,Aaboud:2018aqj}. Unfortunately, these searches are not sufficiently sensitive to the FTH setup. The lighter glueball masses for which the decay length is long enough to produce DV signatures in the HCAL and the muon system are already in tension with constraints from DV searches in light Higgs decays.

For the glueball masses of interest to us, the majority of the DVs occur in the tracker. The only dedicated tracker DV trigger used so far is the $H_T$ trigger used by CMS~\cite{CMS:2014wda,Sirunyan:2018vlw}. In these searches $H_T$ is defined as the scalar sum of the transverse energies of all jets passing certain selection cuts, which are optimized for the decay of heavier particles. Since in the FTH setup, $G_0$ is only one of several possible glueballs in the final state, and due to additional efficiency factors for the DV occurring in the active detector volume, no more than a few events are likely to pass these cuts even at high luminosity.

\subsection{Heavy Higgs to Di-Higgs}\label{ss.HhhDVX}
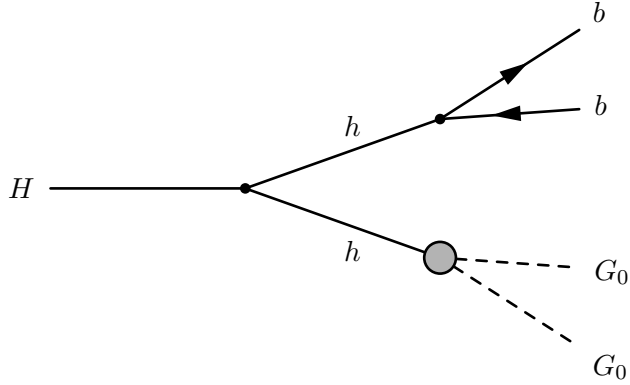
\begin{figure}[t]
\begin{fmffile}{Htohh}
\begin{fmfgraph*}(200,120)
\fmfpen{1.0}
\fmfstraight
\fmfleft{p1,p2,i1,p3,p4} \fmfright{o1,o2,p5,o3,o4}
\fmfv{l= $H$}{i1}
\fmf{plain,tension=2}{i1,v1}
\fmf{plain,tension=1}{v1,v2}\fmf{plain,tension=1}{v1,v3}
\fmf{dashes,tension=0.7}{v2,o2}\fmf{dashes,tension=0.7}{v2,o1}
\fmf{fermion,tension=0.7}{o3,v3,o4}
\fmfv{decor.shape=circle,decor.filled=full,decor.size=1.5thick}{v1}
\fmfv{decor.shape=circle,decor.filled=full,decor.size=1.5thick,l=$h$,l.a=185,l.d=30}{v3}
\fmfv{decor.shape=circle,decor.filled=30,decor.size=6thick,l=$h$,l.a=175,l.d=30}{v2}
\fmfv{l=$G_0$}{o1}\fmfv{l=$G_0$}{o2}
\fmfv{l=$b$}{o3}\fmfv{l=$b$}{o4}
\end{fmfgraph*}
\end{fmffile}
\caption{\label{f.Htohh} Diagram for the $H\to hh\to$~DV+X channel. One light Higgs decays visibly while the other decays to glueballs, leading to a DV, visible objects and $E_T^{\text{mis}}$ for triggering in the event.}
\end{figure}

As we now show, the $H\to hh\to$~DV+X is a more promising discovery channel in the FTH setup. The heavy Higgs branching fraction is $\text{BR}(H\to hh)\gtrsim 1/7$, while the light Higgs branching fraction, suppressed by the small mixing between it and the glueball states, varies between $10^{-2}$ to $10^{-5}$ as the twin top mass varies between 400 to 1000 GeV. This leaves enough room for one light Higgs to decay to glueball states, leading to a DV as well as missing transverse energy ($E_T^\text{mis}$), while the other decays visibly, facilitating triggering. This is shown in Fig.~\ref{f.Htohh}. This final state can satisfy the search criteria of the ATLAS study requiring only one DV~\cite{Aaboud:2017iio}. While backgrounds are larger for a single DV in the final state, the search of ref.~\cite{Aaboud:2017iio} requires a large ($n_\text{Tracks}\geq 5$) number of tracks emerging from the DV with an invariant mass above 10~GeV, reducing the expected background to 0.02 events at 32.8 $\text{fb}^{-1}$ at 13 TeV.

We estimate the signal acceptance for this search in the FTH model using Monte Carlo simulation. In particular, we simulate the kinematics of heavy twin Higgs production in {\sc MadGraph}~\cite{Alwall:2014hca}, and we take the production cross section values from the Higgs Cross Section Working Group~\cite{deFlorian:2016spz}, modifying them according to Eq.~\eqref{e.production} for the FTH model. The decays of $H\to hh$, $h\to b\bar{b}$ and $h\to G_0G_0$ are isotropic. The $b$-jets must pass preselection cuts $p_T>25$ GeV and $|\eta|<3$. The DV preselection cuts are summarized in Table~\ref{t.cuts}.

\begin{table}
\begin{center}
\begin{tabular}{| l || c |}
\hline
Prompt $b$-jets & $p_T>$25 GeV, $|\eta|<3$ \\
\hline 
DV & $n_\text{Tracks}\geq5$ with ${p_T}_\text{Track}>$1 GeV, $m_\text{DV}>10$ GeV
 \\
\hline
Event & $E_T^\text{mis}>$ 130 GeV
\\
\hline
\end{tabular}
\end{center}
\caption{
Preselction cuts for displaced vertices in the search of~\cite{Aaboud:2017iio}. In final selection the $E_T^\text{mis}$ cut is increased to 250~GeV.
}
\label{t.cuts}
\end{table}

For the $G_{0}$ decay, we focus our attention on the $b\overline{b}$ final state, taking the associated branching fraction into account. The decay products of the $b$'s generically provide sufficiently many tracks to satisfy the $n_\text{Tracks}\geq 5$ and the $m_\text{DV}>10$ GeV requirements. We use {\sc{FeynRules}}~\cite{Alloul:2013bka} and {\sc MadGraph} interfaced with {\sc Pythia}8~\cite{Sjostrand:2014zea} to simulate the $G_{0}$ decays, reading off the final state tracks from the {\sc Pythia} event record. After boosting the decaying $G_{0}$ particles according to the kinematic distributions described above, we impose the selection cuts of Table~\ref{t.cuts}. The $E_T^\text{mis}\geq$250 GeV requirement can be satisfied, if only inefficiently, due to the second glueball that escapes the detector and carries away invisible momentum.

\begin{figure}[th]
\begin{centering}
\includegraphics[width=0.49\textwidth]{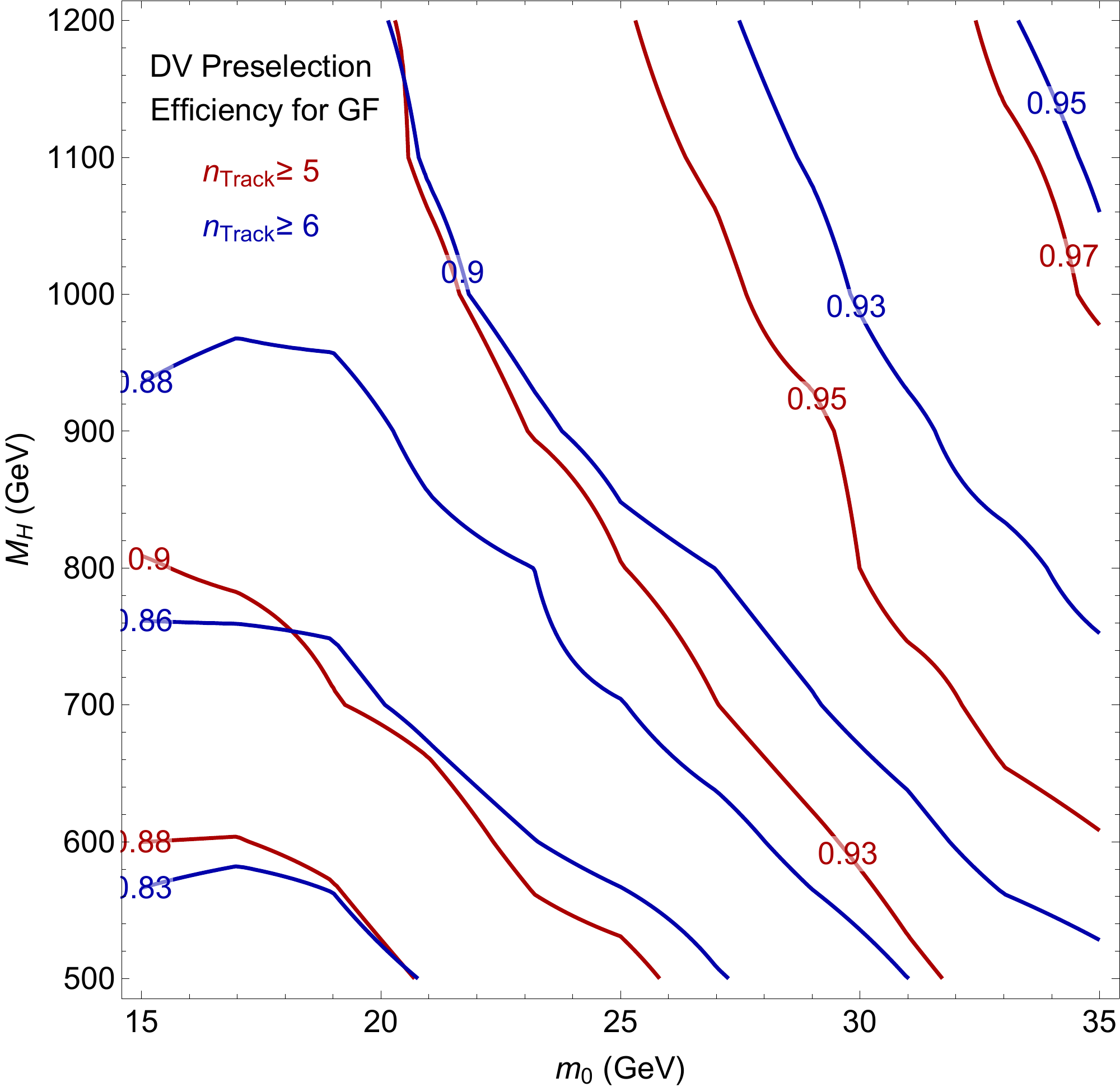}
\includegraphics[width=0.49\textwidth]{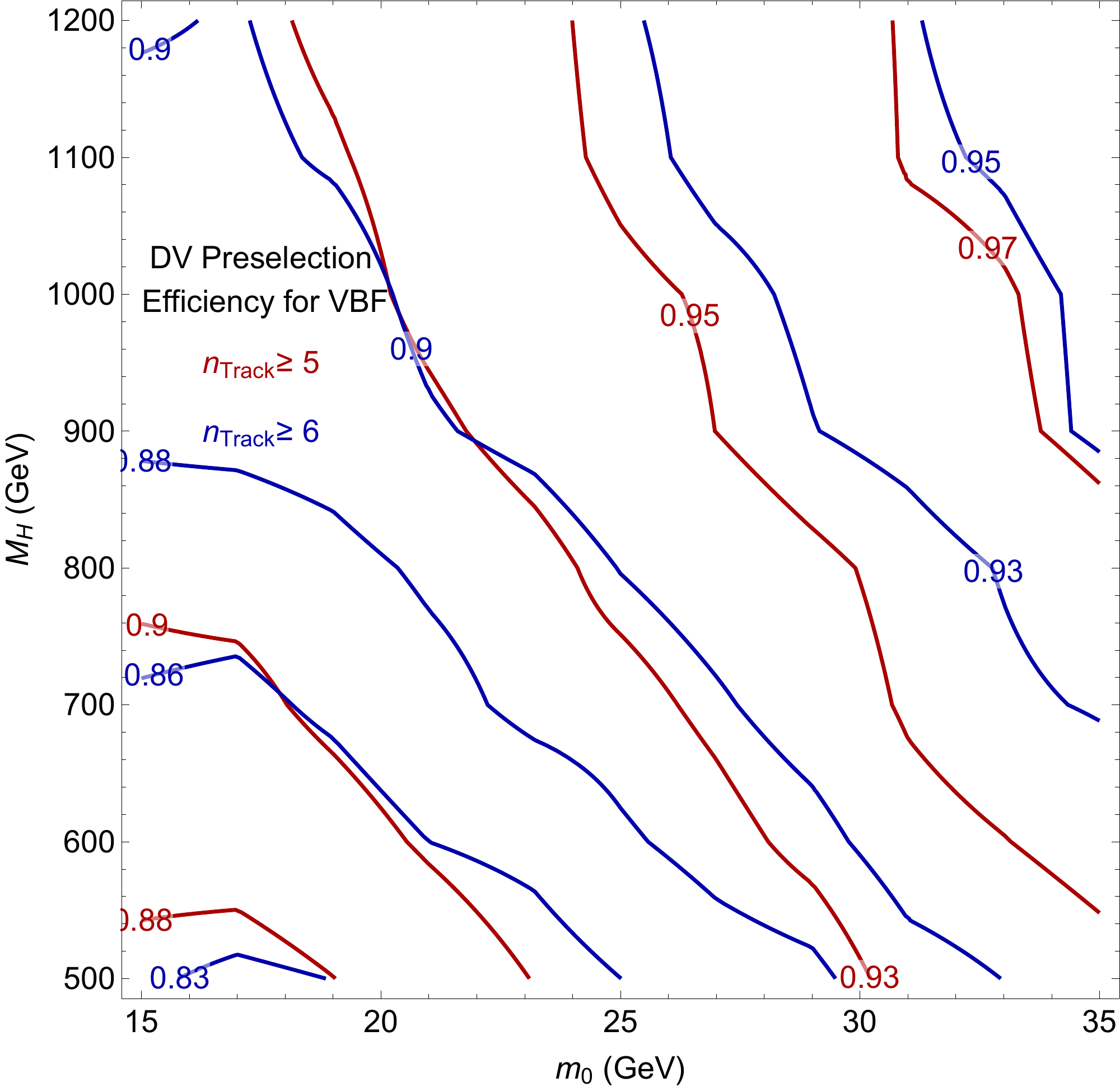}
\end{centering}
\caption{\label{f.DVeff} Contours of efficiency (red) for passing the DV preselection cuts summarized in Table~\ref{t.cuts} for a heavy twin Higgs produced through gluon fusion (VBF) on the left (right). For the blue contours, the $n_\text{Tracks}$ requirement is increased from 5 to 6.}
\end{figure}

In Fig.~\ref{f.DVeff} we show the contours (in red) of the fraction of events where the DV takes place in the tracker volume specified in~\cite{Aaboud:2017iio}, and where the tracks meet the selection requirements as a function of $m_H$ and $m_0$. For nearly the entire parameter region, this fraction is above $\gtrsim$90\% for production through gluon fusion (left) and VBF (right). In plotting the blue contours, the $n_\text{Tracks}$ requirement is increased from 5 to 6, which according to~\cite{Aaboud:2017iio} leads to an additional background reduction by more than 50\%. A potential concern for the FTH setup is a reduction in the DV reconstruction efficiency due to the displaced nature of $b$-meson decays. However, this is expected to be a small effect for the b-meson boost factors corresponding to the range of glueball mass we consider. 

\begin{figure}[th]
\begin{centering}
\includegraphics[width=0.325\textwidth]{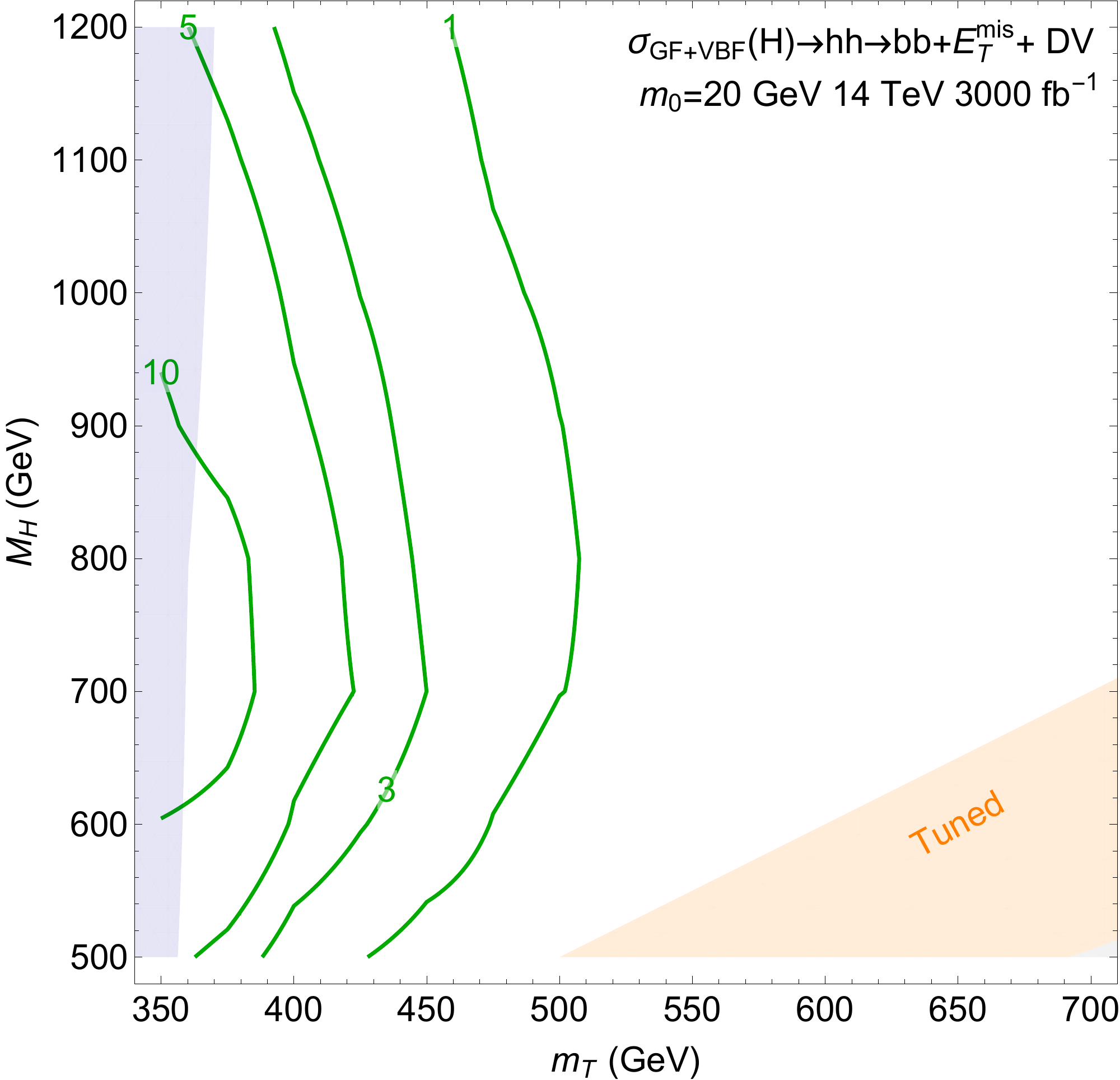}
\includegraphics[width=0.325\textwidth]{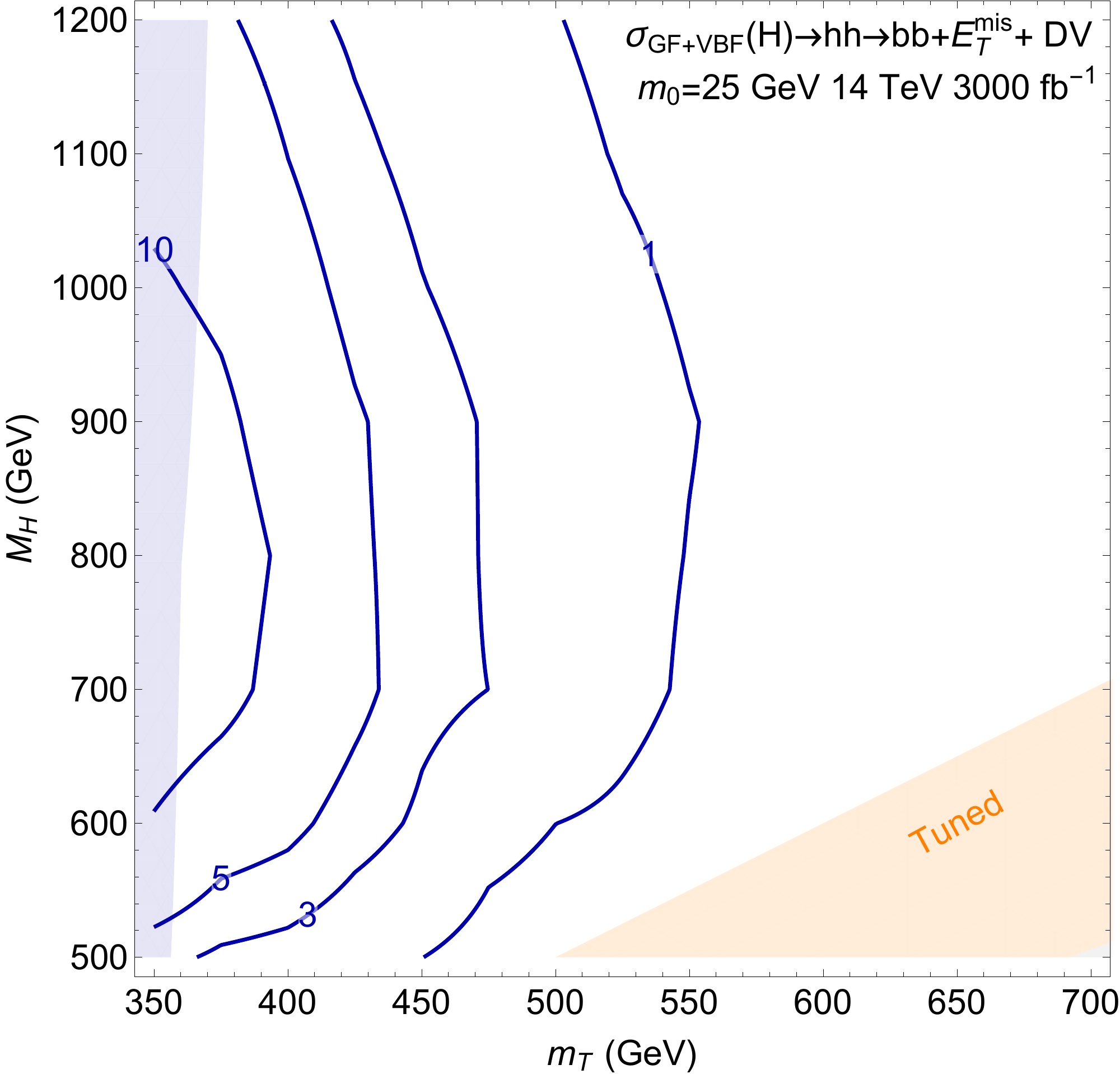}
\includegraphics[width=0.325\textwidth]{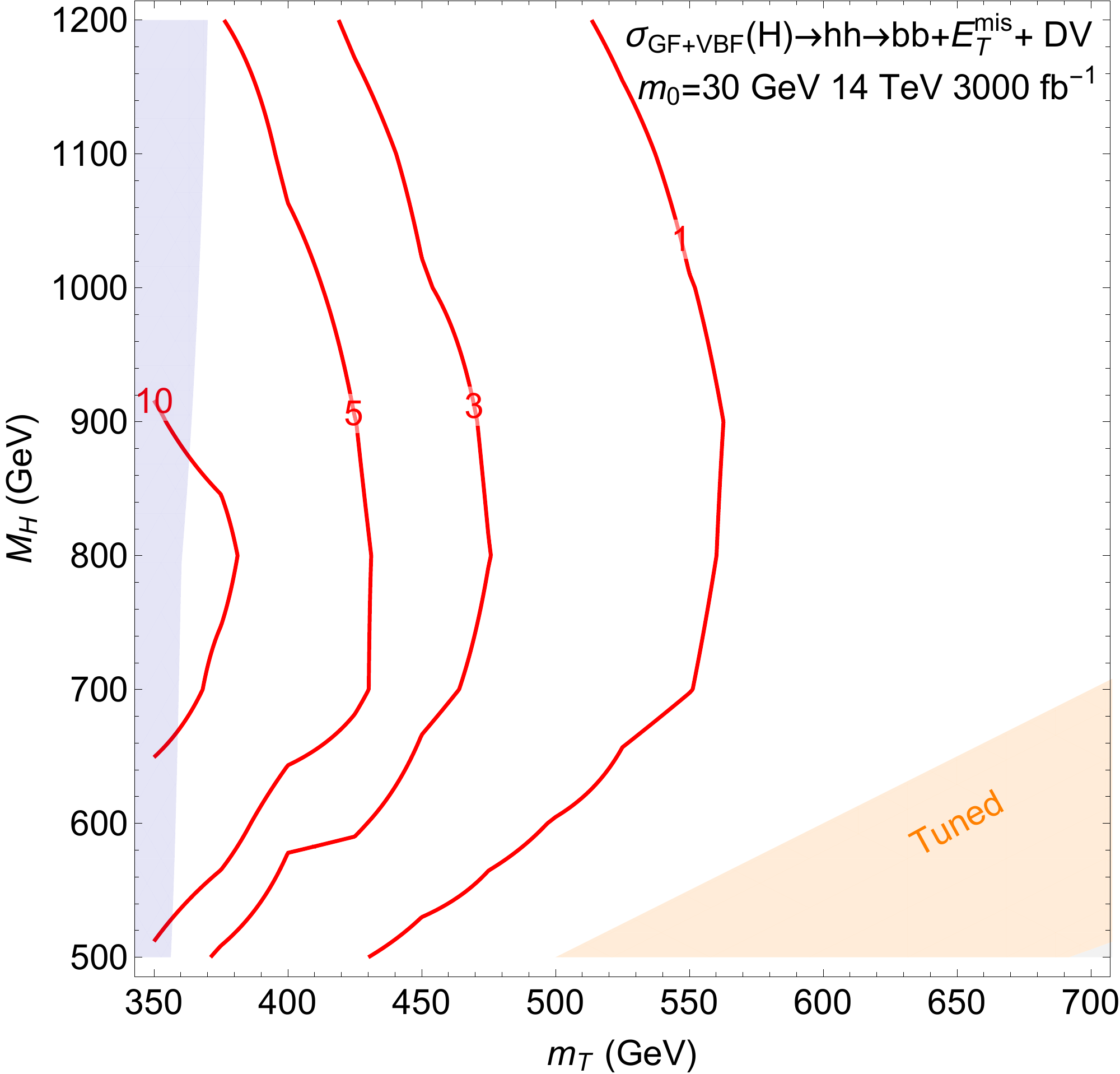}
\end{centering}
\caption{\label{f.HtohhResults} Contours of the number of the signal event count after all acceptance and efficiency factors have been taken into account, with 3000 $\text{fb}^{-1}$ of luminosity at the HL-LHC (14 TeV). In the left, middle, and right plots, the mass of the $G_0$ state is taken to be 20, 25, and 30 GeV respectively. In the orange shaded region, the tuning is not improved over the SM, while the blue shaded region is in tension with Higgs coupling measurements. }
\end{figure}

When calculating the final signal rate, in addition to the geometric acceptance plotted in Fig.~\ref{f.DVeff}, we also factor into our analysis the DV reconstruction efficiency given in Fig.~2b of~\cite{Aaboud:2017iio}, and we add an additional multiplicative factor of $0.55$ corresponding to the fraction of the tracker volume that can be used for DV reconstruction. In Fig.~\ref{f.HtohhResults} we plot contours of the expected number of signal events at the HL-LHC, assuming 14 TeV collisions and 3000 $\text{fb}^{-1}$ of luminosity, for three glueball masses: 20, 25, and 30 GeV. The ATLAS collaboration has studied how this search can be applied to the HL-LHC~\cite{ATL-PHYS-PUB-2018-033}, with two options for the background rate extrapolation: a linear scaling with the luminosity which yields $1.8^{+1.8}_{-0.9}$ events, or a more optimistic scenario with a total expected background of $0.02^{0.02}_{-0.01}$. In either case, we take a simplistic approach and we consider an expected signal event count of 5 to be sufficient for exclusion, and a signal count of 10 to be sufficient for discovery.

We see from Fig.~\ref{f.HtohhResults} that the heavy Higgs can be discovered for smaller $m_T$ and heavier $m_0$. Recall from Sec.~\ref{ss.lightHiggstToGlue} that DV's arising from light Higgs decays probe a complementary region of parameter space. Therefore, the parameter space can be explored more completely by leveraging both searches. These searches are most powerful at higher $m_H$, where the glueball escaping the detector carries away more energy, but this increase in efficiency is eventually counteracted by the reduction in the $H$ production cross section. 

\begin{figure}[th]
\begin{centering}
\includegraphics[width=0.325\textwidth]{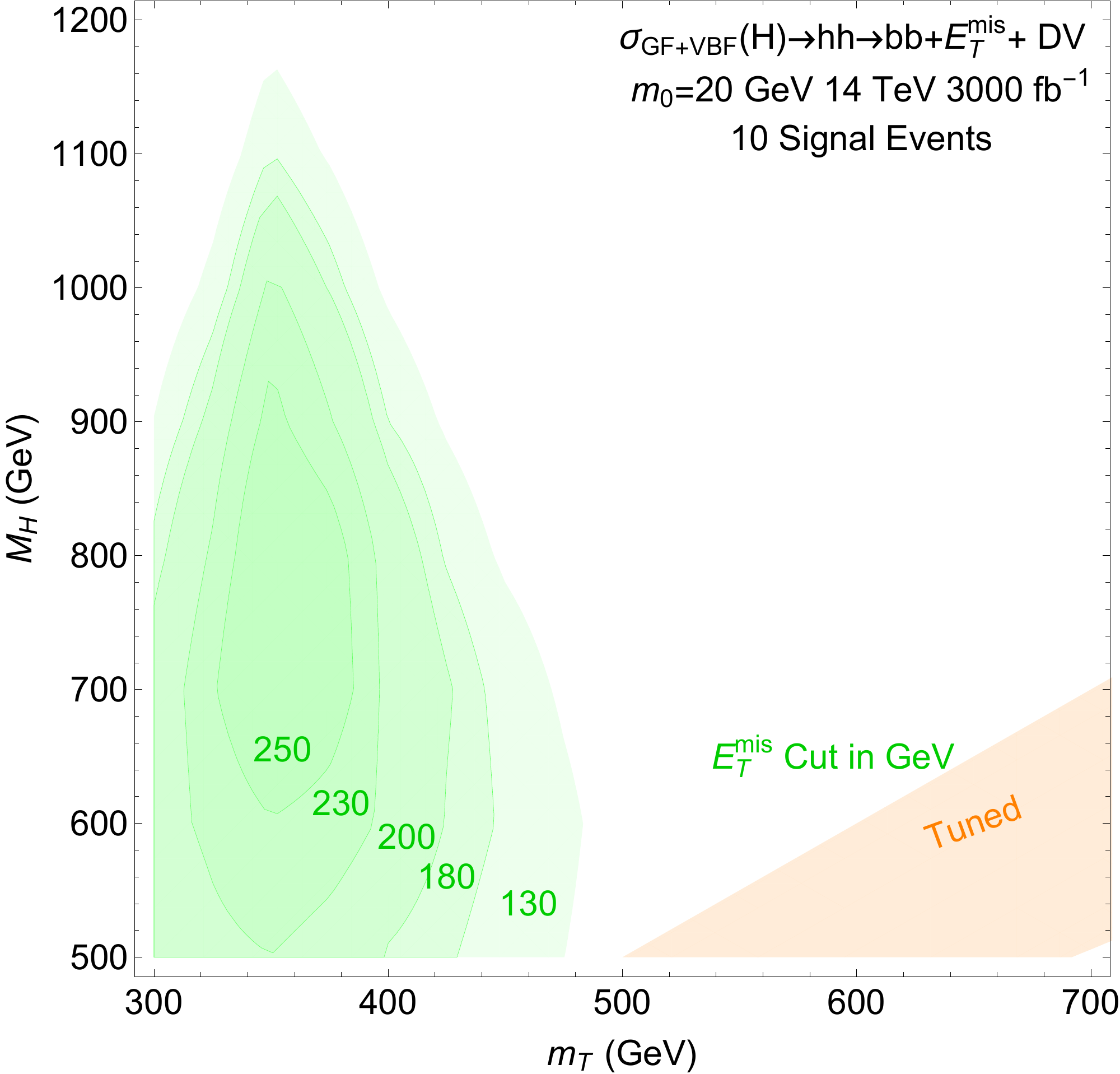}
\includegraphics[width=0.325\textwidth]{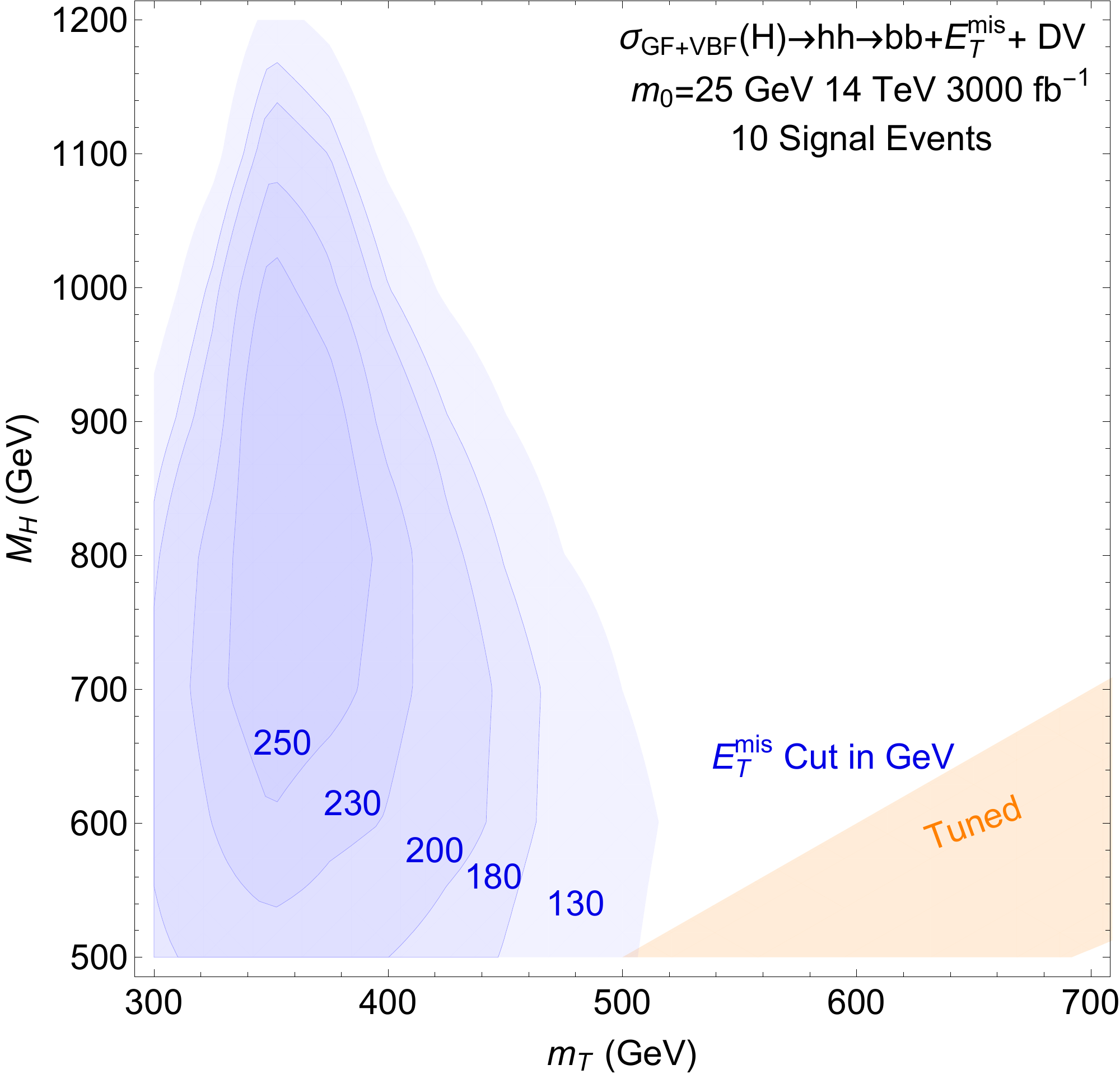}
\includegraphics[width=0.325\textwidth]{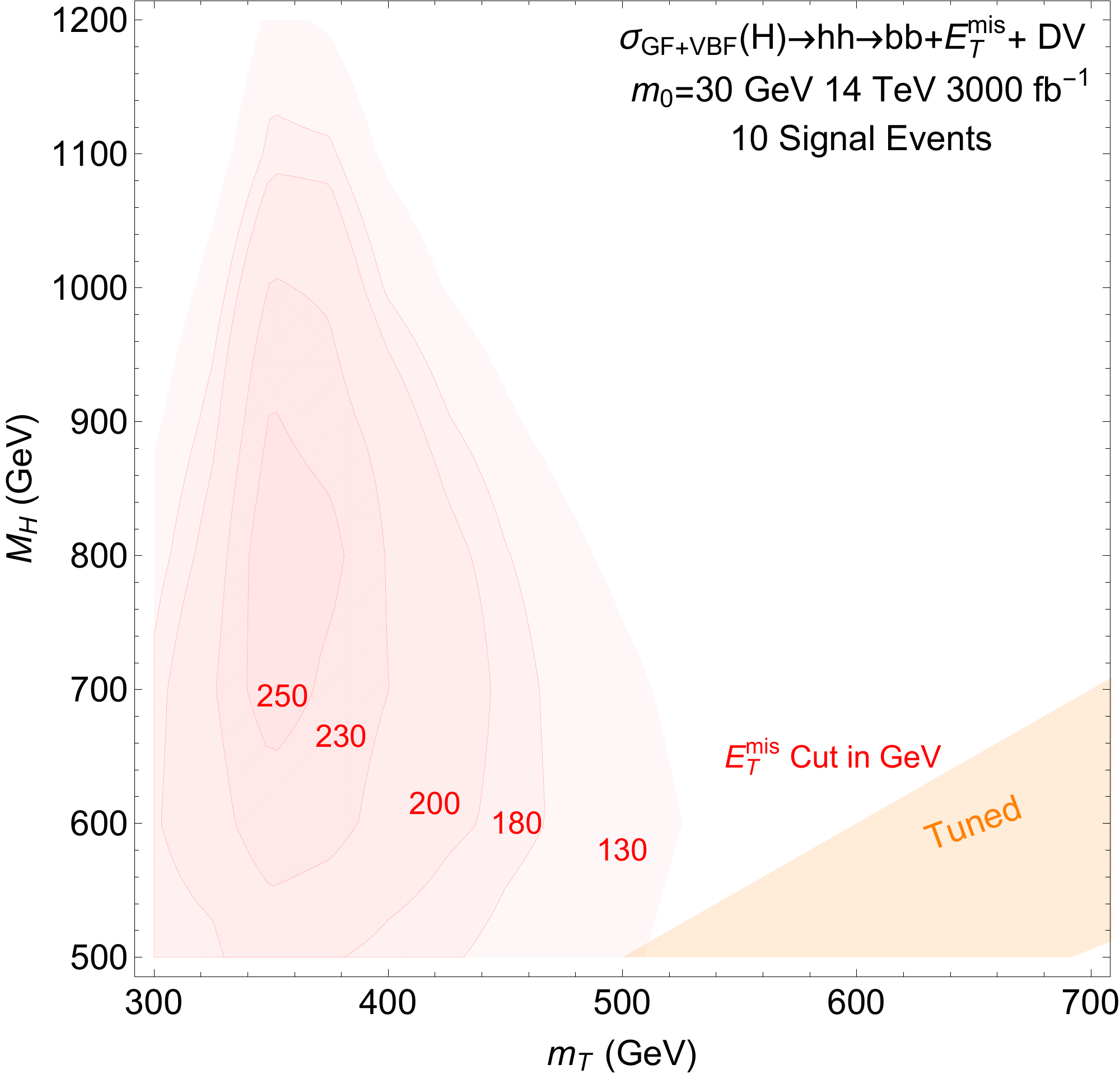}
\end{centering}
\caption{\label{f.varyMET}Regions with more than ten signal events with 3000 $\text{fb}^{-1}$ of luminosity at the HL-LHC (14 TeV) for a range of $E^\text{mis}_{T}$ cut thresholds. In the left, middle, and right plots, the mass of the $G_0$ state is taken to be 20, 25, and 30 GeV respectively. In the orange shaded region, the tuning is not improved over the SM.}
\end{figure}

The escaping glueball typically has enough transverse energy to satisfy the preselection cut $E^\text{mis}_{T}>130$ GeV, while the full selection cut of $E^\text{mis}_{T}>250$ GeV presents a significant challenge, requiring larger $H$ masses and hence lower production cross sections. In Fig.~\ref{f.varyMET} we show how relaxing the $E^\text{mis}_{T}$ cut changes the discovery contour (for 10 signal events after all cuts) for the three glueball mass benchmarks. The outermost region corresponds to the preselection cut, and the innermost to the final selection cut used in~\cite{Aaboud:2017iio}. We see that reducing the cut threshold even mildly can substantially increase the discovery reach. Of course, relaxing the  $E^\text{mis}_{T}$ cut will increase the background rate. However, since the cut value used in~\cite{Aaboud:2017iio} was optimized for a very different signal model, a lower cut may well yield an enhanced discovery reach for the FTH scenario. As we saw in Fig.~\ref{f.DVeff}, relaxing the $E^\text{mis}_{T}$ cut can also be combined with a more stringent cut on the number of tracks emerging from the DV for reducing the background, without a significant reduction in the signal rate.

\subsection{Exploring the Twin Sector}
The displaced glueball decays offer more than a new discovery channel for the heavy Higgs boson. They can also be used to measure parameters of the underlying microscopic physics, and test the self-consistency of the Twin Higgs mechanism. In ref.~\cite{Chacko:2017xpd} a similar strategy was proposed that relies on extracting parameters in the scalar potential from prompt decays of the heavy Higgs. The discovery of displaced glueball decays would make it possible to extend the consistency check to the twin fermion sector.

In particular, measuring the glueball mass and lifetime allows one to fit two otherwise undetermined model input parameters. The glueball mass is proportional to the twin QCD scale, which for simplicity we take to be equivalent to the Landau pole in the one-loop running of $\alpha_{s}^\text{B}$. Assuming identical initial conditions for $\alpha_{s}^\text{A}$ and $\alpha_{s}^\text{B}$ at $\Lambda_\text{UV}$, the Landau pole is in turn determined by the twin quark masses, with the twin top mass related to the SM top mass by a factor of $\tan\vartheta$. If measurements similar to those described in ref~\cite{Chacko:2017xpd} are performed at the HL-LHC that determine the relevant parameters in the twin scalar sector, then a measurement of the glueball mass adds an indirect probe of the UV completion scale of the Twin Higgs model. Furthermore, with the glueball mass and the parameters in the twin scalar sector known, Eq.~\ref{e.glueballwidth} can be used to predict the lifetime of the glueball, which can be tested against the experimental measurement of $c\tau$, providing a nontrivial check of the ${\mathbb Z}_{2}$ symmetry structure in the underlying model. The full symmetry structure of the Twin Higgs setup is defined not only by the ${\mathbb Z}_{2}$ symmetry, but also the approximate $SU(4)$ global symmetry of the scalar sector. This latter symmetry plays an essential role in setting the $H\to hh$ branching ratio, and therefore a comparison of the expected and observed signal event rates combined with the measurement of the glueball mass and lifetime can be used to probe not only the existence of the ${\mathbb Z}_{2}$ symmetry but the full symmetry structure of the Twin Higgs framework.

There are a number of challenges associated with this proposal. As we have seen, the number of signal events is generically not large, unless the $E^\text{mis}_{T}$ cut is lowered to below 250~GeV, and a low signal event count would introduce significant statistical uncertainty into the measurements of both the glueball mass and its lifetime. The glueball mass measurement also suffers from detector (specifically, HCAL) energy resolution, unless a sophisticated fitting procedure can be devised that relies on track information alone. The prediction of the glueball mass in terms of the twin QCD scale, as well as the matrix elements appearing in the glueball lifetime prediction have theoretical uncertainties which need to be propagated through the calculations described above. Nevertheless, the measurement of the glueball mass and lifetime appears intriguing in terms of providing an indirect probe into the twin fermion sector of the theory which is otherwise experimentally inaccessible. It is important to realize that a similar procedure applies to the displaced decays of the light Higgs, and indeed the possibility of more DVs in such an analysis may improve the prospects of success. Finally, the prediction of the signal event rate that is necessary for probing the full symmetry structure of the Twin Higgs mechanism depends on additional order one factors arising from twin hadronic matrix elements that may need to be calculated more precisely, possibly through lattice studies.

\section{Conclusion\label{s.con}}
\begin{figure}[t]
\begin{centering}
\includegraphics[width=0.8\textwidth]{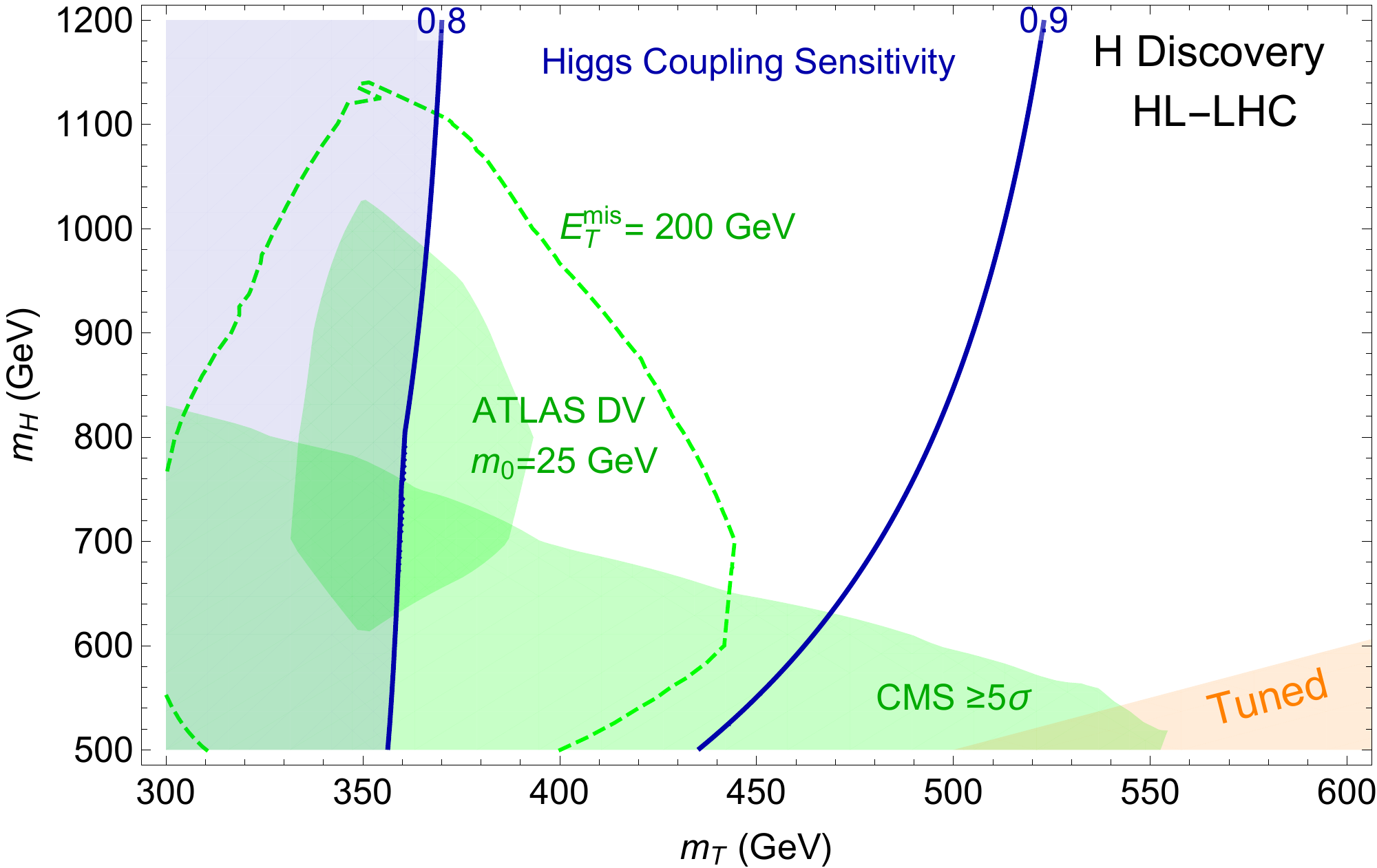}
\end{centering}
\caption{\label{f.final} Discovery prospects for the heavy Higgs in the FTH setup at HL-LHC with 3000 $\text{fb}^{-1}$ of luminosity at the HL-LHC (14 TeV). The CMS reach is based on the prompt decay $H\to ZZ\to 4\ell$, while the ATLAS reach is based on the $H\to hh\to$~DV+X channel described in section~\ref{ss.HhhDVX}. The dashed contour shows the increase in the parameter region where 10 signal events pass all cuts when the $E_T^\text{miss}$ cut is reduced to 200 GeV. The blue contours indicate the ratio of $(\sigma\times{\rm BR})$ of the light Higgs to that of the SM Higgs.}
\end{figure}

We have investigated the prospects for discovering the heavy Higgs in the FTH framework through displaced decays of glueball states. While glueballs can be produced by the direct decay of the heavy Higgs into twin gluons, this presents a challenge for triggering. However, the decay channel $H\to hh\to$~DV+X is more promising. When one light Higgs decays into twin glueballs, one of which decays in the tracker, selection cuts such as those used in~\cite{Aaboud:2017iio} can be satisfied, making discovery possible. In addition, the region where a DV-based search is sensitive is complementary to channels based on prompt decays of the heavy Higgs, and displaced decays of the light Higgs. The reach in these complementary channels is illustrated In Fig.~\ref{f.final}, with the DV-based search extending the HL-LHC's heavy Higgs discovery potential to larger values of $m_H$. The dramatic effect of reducing the $E^\text{mis}_{T}$ cut to 200 GeV is also shown, with the sensitivity improving at still higher values of $m_H$, as well as higher values of $m_T$. The DV-based discovery region is contained within the region which will be probed though Higgs coupling measurements indicated by the blue contours. Therefore, the complementary information provided by the Higgs coupling measurements and a discovery in the DV-based search can be combined to test the twin Higgs framework, and probe the twin fermion sector.

\acknowledgements
We are pleased to thank Zackaria Chacko, David Curtin, and Ennio Salvioni for advice and discussion. We also thank Laura Jeanty and Christian Ohm for their help understanding the ATLAS searches. The research of C.K. is supported by the National Science Foundation Grant Number PHY-1620610. S.N. is supported by Vrije Universiteit Brussel through the Strategic Research Program ``High Energy Physics" and also supported by FWO under the EOS-be.h project n. 30820817. C.B.V. is supported by Department of Energy Grant Number DE-SC-000999. 

\appendix

\section{Model Details\label{a.Model}}
In this Appendix we provide some of the details of the twin Higgs set up, see also~\cite{Chacko:2017xpd}. We begin by determining the potential parameters of Eq.~\eqref{e.HiggsPotential} in terms of the mass eigenstates:
\begin{align}
\lambda^2&=\frac{1}{16f^4}\left[(m_{H}^2-m_{h}^2)^2-4\cot^2(2\vartheta)m_{H}^2m_{h}^2 \right],\\
\delta&=\frac{1}{4f^2}\left[(m_{h}^2+m_{H}^2)-\sqrt{(m_{H}^2-m_{h}^2)^2- 4 m_{h}^2 m_{H}^2\cot^2(2\vartheta)} \right].
\end{align}
Note that in order to keep $\lambda^2\geq 0$ we must have
\begin{equation}
\frac{m_{H}}{m_{h}}\geq |\cot(2\vartheta)|+|\csc(2\vartheta)|.
\end{equation}
As we associate the $m_{h}$ with the observed 125 GeV Higgs and expect $\vartheta<\pi/4$, this gives a lower bound on $m_{H}$ as a function of $v/f$:
\begin{equation}
m_{H}\geq m_{h}\cot\vartheta=m_h\frac{m_T}{m_t}.
\end{equation}

The kinetic term for $\mathcal{H}$ leads to 
\begin{align}
&\frac12\partial_\mu\sigma\partial^\mu\sigma+\frac12\left( 1+\frac{\sigma}{\sqrt{2}f}\right)^2\partial_\mu \rho\partial^\mu \rho\nonumber\\
&+\left[\frac{f^2g^2}{2}W_{A\mu}^{+}W_A^{\mu-}+\frac{f^2g^2}{4\cos^2\theta_{W}} Z_{A\mu}Z_A^{\mu} \right]\left( 1+\frac{\sigma}{\sqrt{2}f}\right)^2\sin^2\left(\frac{v+\rho}{\sqrt{2}f} \right)\nonumber\\
&+\left[\frac{f^2g^2}{2}W_{B\mu}^{+}W_B^{\mu-}+\frac{f^2g^2}{4\cos^2\theta_{W}} Z_{B\mu}Z_B^{\mu} \right]\left( 1+\frac{\sigma}{\sqrt{2}f}\right)^2\cos^2\left(\frac{v+\rho}{\sqrt{2}f} \right).
\end{align}
From which we find
\begin{equation}
\begin{array}{cc}
\displaystyle M^2_{W_A}=\frac{f^2g^2}{2}\sin^2\vartheta, & \displaystyle M^2_{W_B}=\frac{f^2g^2}{2}\cos^2\vartheta, 
\end{array}
\end{equation}
with the $Z$ boson masses related by the usual factor of $\cos\theta_W$. In the mass basis we have
\begin{align}
\left( 1+\frac{\sigma}{\sqrt{2}f}\right)^2\sin^2\left(\frac{v+\rho}{\sqrt{2}f} \right)&=\sin^2\vartheta\left[1+\frac{2h}{v_\text{EW}}\cos(\vartheta-\theta)+\frac{2H}{v_\text{EW}}\sin(\vartheta-\theta)+\ldots\right],\\
\left( 1+\frac{\sigma}{\sqrt{2}f}\right)^2\cos^2\left(\frac{v+\rho}{\sqrt{2}f} \right)&=\cos^2\vartheta\left[1-\frac{2h}{v_\text{Twin}}\sin(\vartheta-\theta)+\frac{2H}{v_\text{Twin}}\cos(\vartheta-\theta)+\ldots\right],
\end{align}
which determines the Higgs couplings.

We now turn to the cubic $Hhh$ coupling. The Higgs potential Eq. \eqref{e.HiggsPotential} includes
\begin{equation}
\sigma^3\sqrt{2}f\left[\lambda+\delta-\frac12\delta\sin^2(2\vartheta) \right]-\sigma^2h\frac{5\delta f}{2\sqrt{2}}\sin(4\vartheta) -\sigma h^2\sqrt{2}\delta f\left[1-3\sin^2(2\vartheta) \right]+h^3\frac{\delta f}{\sqrt{2}}\sin(4\vartheta).
\end{equation}
This leads to the cubic term
\begin{align}
g_{Hhh}\equiv&\sqrt{2}f\left\{3(\lambda+\delta)\sin\theta\sin(2\theta)-\frac{\delta}{8}\left[ \cos\theta-9\cos(3\theta)+2\cos(\theta-4\vartheta)\right.\right.\nonumber\\
&\left.\phantom{\frac{2f}{\sqrt{2}}}\left.+\cos(\theta+4\vartheta)+21\cos(3\theta-4\vartheta)\right]\right\}.
\end{align}
The kinetic term adds  
\begin{equation}
\frac{\cos\theta}{\sqrt{2}f}\left[\cos^2\theta H\partial_\mu h \partial^\mu h-2\sin^2\theta h\partial_\mu H \partial^\mu h \right].
\end{equation}
Then the width of $H$ into $h$ pairs is
\begin{align}
\Gamma(H\to hh)=\frac{1}{32\pi m_H}\sqrt{1-4\frac{m_{h}^2}{m_H^2}}&\left[ g_{Hhh}+\frac{2\cos\theta}{\sqrt{2}f}(1+\sin^2\theta)\left(\frac{m_{H}^2}{2}-m_{h}^2 \right)\right.\nonumber\\
&\left.\phantom{A}+\frac{4m_{h}^2}{\sqrt{2}f}\cos\theta\sin^2\theta \right]^2.
\end{align}

Finally we consider the top-quark sector. All other fermions couplings are similarly defined.
\begin{align}
\lambda_t\left[fq_At_A\left( 1+\frac{\sigma}{\sqrt{2}f}\right)\sin\left(\frac{v+\rho}{\sqrt{2}f} \right) +fq_Bt_B\left( 1+\frac{\sigma}{\sqrt{2}f}\right)\cos\left( \frac{v+\rho}{\sqrt{2}f}\right) \right].\label{e.fermionCoupling}
\end{align}
The mass basis couplings follow from
\begin{align}
\left( 1+\frac{\sigma}{\sqrt{2}f}\right)\sin\left(\frac{v+\rho}{\sqrt{2}f} \right)&=\sin\vartheta\left[1+\frac{h}{v_\text{EW}}\cos(\vartheta-\theta)+\frac{H}{v_\text{EW}}\sin(\vartheta-\theta)+\ldots\right],\\
\left( 1+\frac{\sigma}{\sqrt{2}f}\right)\cos\left(\frac{v+\rho}{\sqrt{2}f} \right)&=\cos\vartheta \left[1-\frac{h}{v_\text{Twin}}\sin(\vartheta-\theta)+\frac{H}{v_\text{Twin}}\cos(\vartheta-\theta)+\ldots\right]\; 
\end{align}

\bibliography{DisplacedTwinHiggsbib}

\end{document}